\documentclass[twocolumn,preprintnumbers,amsmath,amssymb]{revtex4}
\usepackage{graphicx}
\usepackage{times}
\usepackage{helvet}
\usepackage{courier}

\newcommand{\eq}[1]{Eq.~(\ref{eq.#1})} 
\newcommand{\eqbare}[1]{(\ref{eq.#1})} 
\newcommand{\fig}[1]{Fig.~\ref{fig.#1}}

\newcommand{\tbl}[1]{Table~\ref{table.#1}}

\newcommand{\sect}[1]{Sec.~\ref{sect.#1}}
\newcommand{\sectlabel}[1]{\label{sect.#1}}
\newcommand{\eqlabel}[1]{\label{eq.#1}}
\newcommand{\figlabel}[1]{\label{fig.#1}}
\newcommand{\tbllabel}[1]{\label{table.#1}}

\newcommand{\voteTotal}{N_{\mbox{\scriptsize vote}}}
\newcommand{\visitRate}{\nu} 
\newcommand{\frontRate}{\nu_{\mbox{\scriptsize f}}} 
\newcommand{\newRate}{\nu_{\mbox{\scriptsize u}}} 
\newcommand{\friendsRate}{\nu_{\mbox{\scriptsize friends}}} 

\newcommand{\fractionToPage}{f_{\mbox{\scriptsize page}}}
\newcommand{\frontPageGrowth}{k_{\mbox{\scriptsize f}}} 
\newcommand{\newPageGrowth}{k_{\mbox{\scriptsize u}}} 

\newcommand{\Votes}{v}			
\newcommand{\fanVotes}{v_F} 	
\newcommand{\nonfanVotes}{v_N} 

\newcommand{\Users}{U}			
\newcommand{\fans}{F} 			
\newcommand{\nonfans}{N}		

\newcommand{\Plognormal}{P_{\mbox{\scriptsize lognormal}}}
\newcommand{\Pprior}{P_{\mbox{\scriptsize prior}}}

\newcommand{\probUserIsAFan}{\rho}

\newcommand{\Rfan}{r_F} 	
\newcommand{\Rnonfan}{r_N} 

\newcommand{\Pfan}{P_F} 	
\newcommand{\Pnonfan}{P_N} 

\newcommand{\Tpromotion}{T_{\mbox{\scriptsize promotion}}}

\newcommand{\erfc}{\mbox{erfc}} 
\newcommand{\rms}{{\sc rms}}

\newcommand{\hour}{\mbox{hr}}

\newcommand{\tFinal}{t_{\mbox{\scriptsize final}}}

\newcommand{\figwidth}{2.8in}

\begin{document}
\title{Using Stochastic Models to Describe and Predict \\Social Dynamics of Web Users}

\author{Kristina Lerman}
       \email{lerman@isi.edu}
\author{Tad Hogg}       
\email{tadhogg@yahoo.com}

\begin{abstract}
Popularity of content in social media is unequally distributed, with some items receiving a disproportionate share of attention from users. Predicting which newly-submitted items will become popular is critically important for both hosts of social media content and its consumers. Accurate and timely prediction would enable hosts to maximize revenue through differential pricing for access to content or ad placement. Prediction would also give consumers an important tool for filtering the ever-growing amount of content.
Predicting popularity of content in social media, however, is challenging due to the complex interactions between content quality and how the social media site chooses to highlight content. Moreover, most social media sites also selectively present content that has been highly rated by similar users, whose similarity is indicated implicitly by their behavior or explicitly by links in a social network.
While these factors make it difficult to predict popularity \emph{a priori}, we show that stochastic models of user behavior on these sites allows predicting popularity based on early user reactions to new content. By incorporating the various mechanisms through which web sites display content,
such models improve on predictions based on simply extrapolating from the early votes.
Using data from one such site, the news aggregator Digg, we show how a stochastic model of user behavior distinguishes the effect of the increased visibility due to the network from how interested users are in the content. We find a wide range of interest, identifying stories primarily of interest to users in the network (``niche interests'') from those of more general interest to the user community. This distinction is useful for predicting a story's eventual popularity from users' early reactions to the story.
\end{abstract}

\maketitle

\section{Introduction}
Success or popularity in social media is not evenly distributed. Instead, a small number of users dominate the activity on the site and receive most of the attention of other users. The popularity of contributed items likewise shows extreme diversity.
For example, relatively few of the four billion images on the social photo-sharing site Flickr are viewed thousands of times, while most of the rest are rarely viewed. Of the tens of thousands of new stories submitted daily to the social news portal Digg, only a handful go on to become wildly popular, gathering thousands of votes, while most of the remaining stories never receive more than a single vote from the submitter herself. Among thousands of new blog posts every day, only a handful become widely read and commented upon.
Given the volume of new content, it is critically important to provide users with tools to help them sift through the vast stream of new content to identify interesting items in a timely manner, or least those items that will prove to be successful or popular.
Accurate and timely prediction will also enable social media companies that host user-generated content to maximize revenue through differential pricing for access to content or ad placement, and encourage greater user loyalty by helping their users quickly find interesting new content.

Success in social media is difficult to predict. Although early and late popularity, which can be measured in terms of
user interest, e.g., votes or views,
an item generates from its inception, are somewhat correlated~\cite{Gomez08,szabo09}, we know little about what drives success. Does success derive mainly from an item's inherent quality~\cite{Agarwal08}, users' response to it~\cite{CraneSornette08}, or some external factors, such as social influence~\cite{Lerman07digg,Lerman07flickr,Lerman08wosn}? In a landmark study, Salganik et al.~\cite{salganik06} addressed this question experimentally by measuring the impact of content quality and social influence on the eventual popularity or success of cultural artifacts. They showed that while quality contributes only weakly to their eventual success, social influence, or knowing about the choices of other people, is responsible for both the inequality and unpredictability of success.
In their experiment, Salganik et al. asked users to rate songs they listened to. The users were assigned to different groups. In the control group (independent condition), users were simply presented with lists of songs. In the other group  (social influence condition), users were also shown how many times each song was downloaded by other users. The social influence condition resulted in large inequality in popularity of songs, measured by the number of times the songs were downloaded. Although a song's quality, as measured by its popularity in the control group, was positively related to its eventual popularity in the social condition group, the variance in popularity at a given quality was very high. This means that two songs of similar quality could end up with vastly different levels of success. Moreover, when users were aware of the choices made by others, popularity was also unpredictable, meaning that on repeating the experiment, the same song could end up with a very different level of popularity.

Although Salganik et al.'s study was limited to a small set of songs created by unknown bands, its conclusions about inequality and unpredictability of popularity appear to apply to cultural artifacts in general and social media production in particular.
While this would appear to prohibit prediction of popularity, we argue that understanding how the collective behavior of Web users emerges from the decisions made by interconnected individuals allows us to predict the popularity of items from the users' early reaction to them. As in previous works~\cite{Lerman07ic,hogg09b,hogg09c}, we use a stochastic modeling framework to mathematically describe the social dynamics of Web users. This framework represents each user and each submitted item as a stochastic process with a few states, e.g., a simple Markov processes whose future state depends only on its present state and the input it receives.
We used this approach to study collective user activity on the social news aggregator Digg. We produced a model that partially explains --- and predicts~\cite{Lerman10www} --- the social voting patterns on Digg and related these aggregate behaviors to the ways Digg enables users to discover new content.  While this model included social influence, i.e., the increased visibility of stories to a user's neighbors in the social network, it did not address the commonality of users' interests indicated by links. This phenomenon, known as \emph{homophily}, is a key aspect of social networks. In this paper we describe a new extension to the model that accounts for systematic variations of interests within and outside of the network.
We make further changes to the model to more closely match it to web site behavior. First, the new model's state transition rates account for the daily variation in user activity~\cite{szabo09}, thereby focusing on variations of votes on individual stories compared to the average activity rate on the site. Second, we account for the variation in number of votes a story receives before it is promoted, which the prior model ignored.

By separating the impact of story quality and social influence on the popularity of stories on Digg, a stochastic model of social dynamics enables two novel applications: (1) estimating inherent story quality from the evolution of its observed popularity, and (2) predicting its eventual popularity based on the early reaction of users to the story. Specifically, to predict how popular a story will become, we can use the early votes, even those cast before the story is promoted, to estimate how interesting it is to voters. With this estimate, the model then determines, on average, the story's subsequent evolution.
We study these claims empirically on a sample of stories retrieved from Digg.
We show that by adjusting for the differing interests among voters, the new model improves predictions of popularity from early reactions of users.

The paper is organized as follows. In Section~\ref{sec:digg} we describe details of the social news aggregator Digg, which provides an empirical foundation and a data set for investigating the utility of stochastic models on the prediction task.  Section~\ref{sec:stochastic} presents an overview of the stochastic modeling framework. In Section~\ref{sect.icwsm_model} we apply the framework to study dynamics of social voting on Digg. We review an existing model of social dynamics of Digg and show that it explains many of the empirically observed features of aggregate behavior of voters on that site. In Section~\ref{sect.niche_model} we extend this model to include variations in story interest to users. Then, in Section~\ref{sec:prediction} we show how  the model can predict eventual popularity of newly submitted stories on Digg.

\section{Social News Portal Digg}
\label{sec:digg}

With over 3 million registered users, the social news aggregator Digg is one of the more popular news portals on the Web.  Digg allows users to submit and rate news stories by voting on, or `digging', them. There are many new submissions every minute, over 16,000 a day. Every day Digg picks about a hundred stories that it believes will be most interesting to the community and promotes them to the front page. Although the exact promotion mechanism is kept secret and changes occasionally, it appears to take into account the number of votes the story receives and how rapidly it receives them. Digg's success is fueled in large part by the emergent front page, which is created by the collective decision of its many users.

While the life cycle of each story may be drastically different from others, its basic elements are the same. These are specified by Digg's user interface, which defines how users can post or discover new stories and interact with other users. A model of social dynamics has to take these elements into account when describing the evolution of story popularity.

\subsection{User interface}
\begin{figure*}[t]
\center{
      \includegraphics[width=0.6\textwidth]{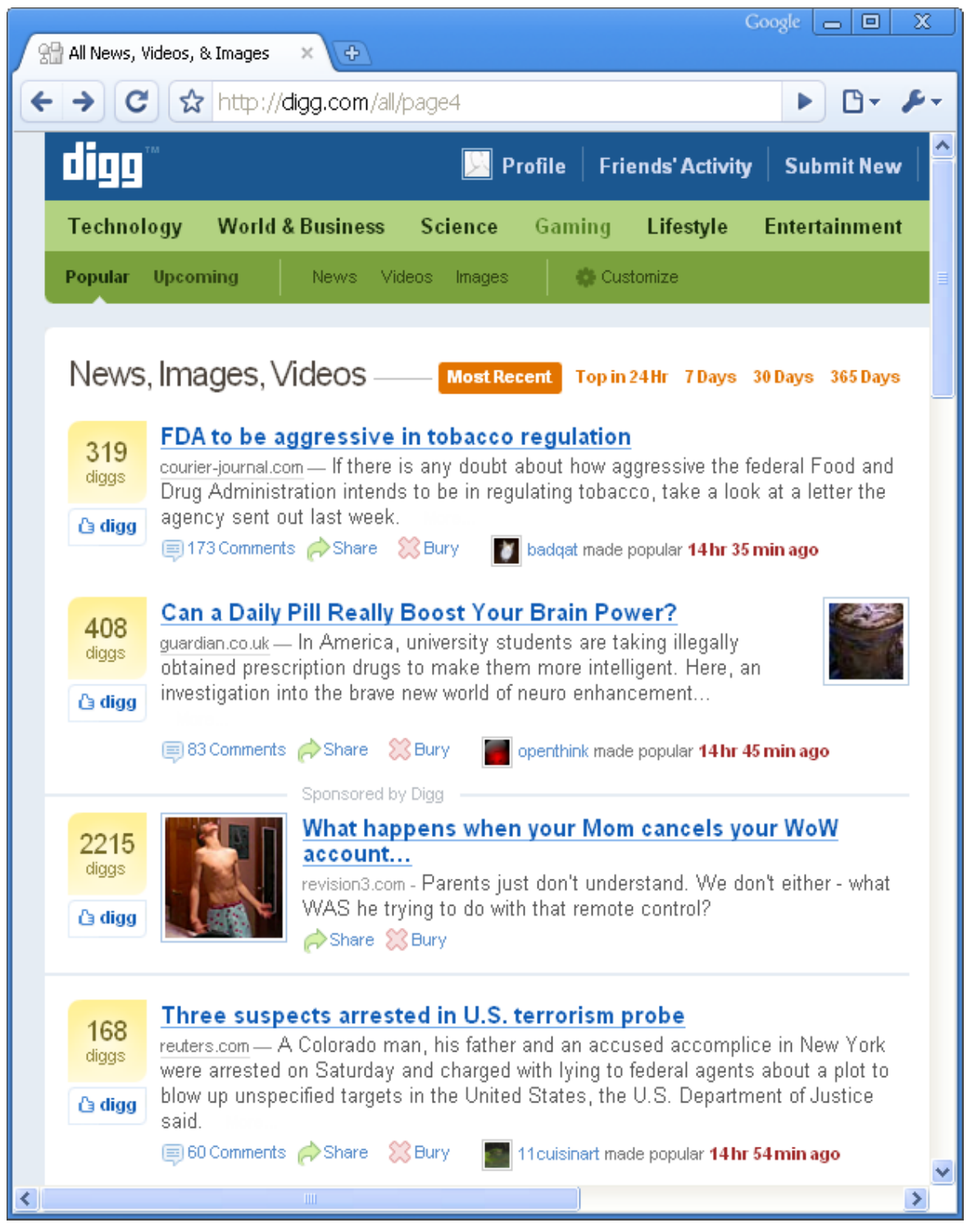} \\
  \caption{Screenshot of the front page of the social news aggregator Digg.}}
  \label{fig:screenshot}
\end{figure*}

A newly submitted story goes on the \emph{upcoming} stories list, where it remains for a period of time, typically 24 hours, or until it is promoted to the front page, whichever comes first.
The default view shows newly submitted stories as a chronologically ordered list, with the most recently submitted story at the top of the list, 15 stories to a page. To see older stories, a user must navigate to page 2, 3, \emph{etc.} of the upcoming stories list. Promoted stories (Digg calls them `popular') are also displayed as a chronologically ordered list on the \emph{front pages}, 15 stories to a page, with the most recently promoted story at the top of the list. To see older promoted stories, user must navigate to page 2, 3, \emph{etc.} of the front page. Figure~\ref{fig:screenshot} shows a screenshot of a Digg front page.
Users vote for the stories they like by `digging' them. The yellow badge to the left of each story  shows its current popularity.

Digg also allows users to designate friends and track their activities, i.e., see the stories friends recently submitted or voted for. The \emph{friends interface} is available through the ``Friends' Activity'' link at the top of any Digg web page (see, for example, Fig.~\ref{fig:screenshot}). The friend relationship is asymmetric. When user $A$ lists user $B$ as a \emph{friend}, $A$ can watch the activities of $B$ but not vice versa. We call $A$ the \emph{fan} of $B$. A newly submitted story is visible in the upcoming stories list, as well as to submitter's fans through the friends interface. With each vote, a story becomes visible to the voter's fans through the friends interface, which shows the newly submitted stories that user's friends voted for.

In addition to these interfaces, Digg also allows users to view the most popular stories from the previous day, week, month, or  year. Digg also implements a social filtering feature which recommends stories, including upcoming stories, that were liked by users with a similar voting history. This interface, however, was not available at the time the data for our study was collected and hence is not part of the stochastic models described in this paper. Thus we examine a period of time where Digg had a relatively simple user interface, which simplifies the stochastic models. This choice allow us to focus on evaluating the  stochastic model approach, particularly the empirical question of whether averaging is useful in the social media setting in spite of long-tail distributions, which contrasts with the narrow distributions found in most statistical physics settings.

\subsection{Dynamics of popularity}

\begin{figure*}[t]
  \begin{tabular}{cc}
      \includegraphics[height=2.2in]{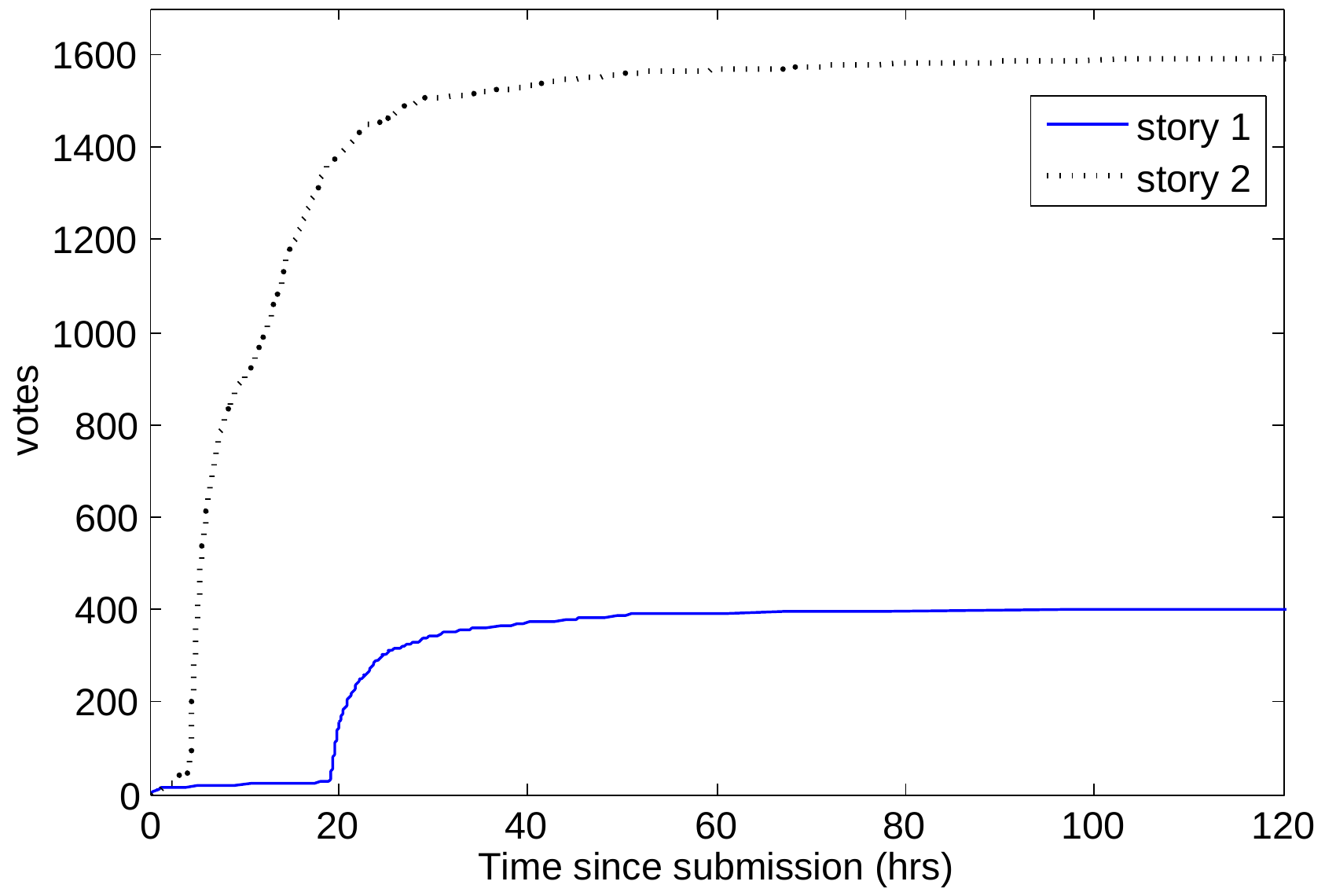} &
      \includegraphics[height=2.2in]{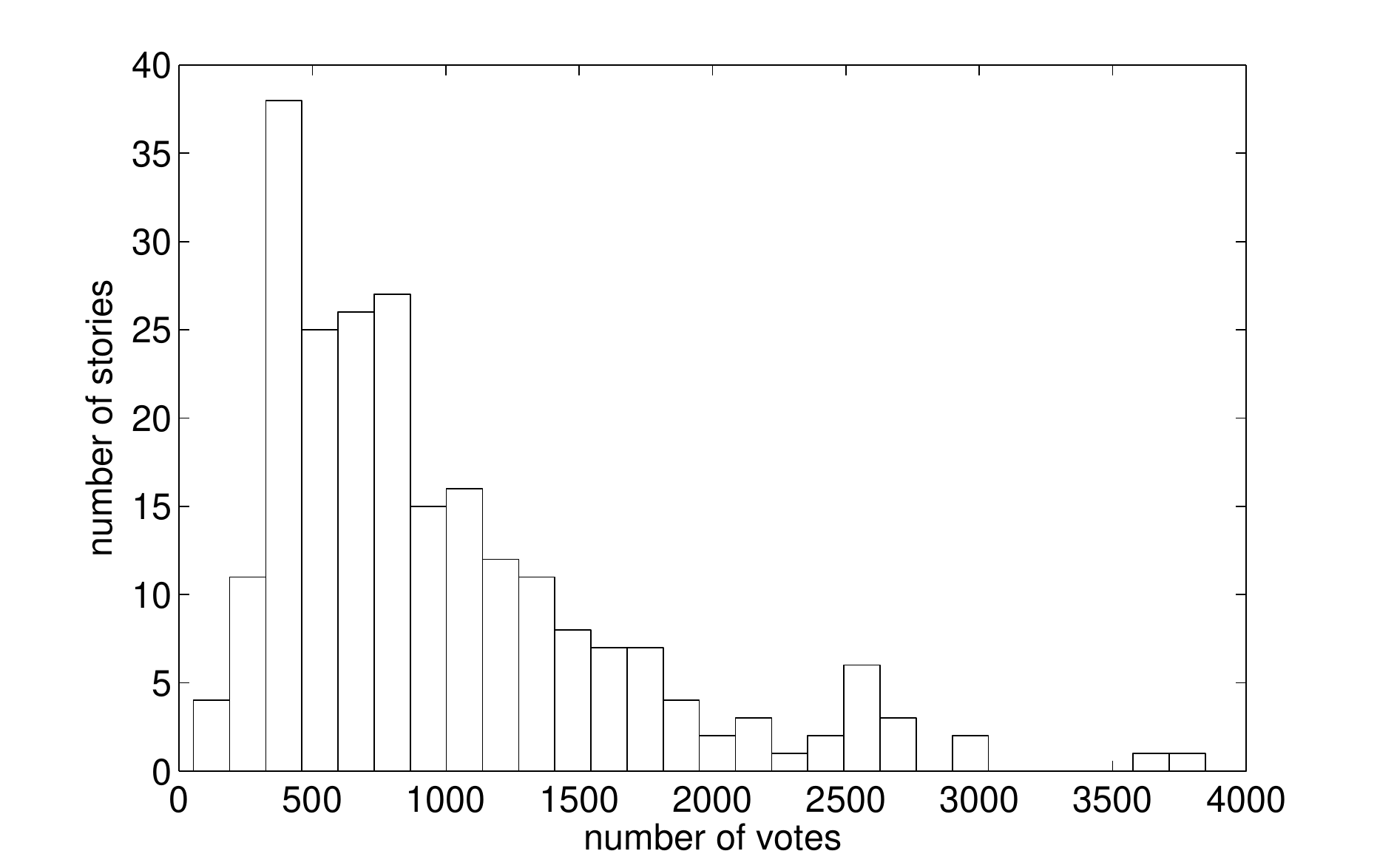}\\
      (a) &        (b)
  \end{tabular}
  \caption{Dynamics of social voting. (a) Evolution of the number of votes received by two front page stories in June 2006. (b) Distribution of popularity of 201 front page stories submitted in June 2006.}\label{fig:votes}
\end{figure*}
While a story is in the upcoming stories list, it accrues votes slowly. If the story is promoted to the front page, it accumulates votes at a much faster pace. Figure~\ref{fig:votes}(a) shows evolution of the number of votes for two stories submitted in June 2006. The point where the slope abruptly increases corresponds to promotion to the front page.
The vast majority of stories are never promoted and, therefore, never experience the sharp rise in the number of votes that accompanies being featured on the front page.
As the story ages, accumulation of new votes slows down~\cite{wu07}, and after a few days the total number of votes received by a story saturates to some value. This value, which we also call the final number of votes, gives a measure of the story's success or {popularity}.

Popularity varies widely from story to story. Figure~\ref{fig:votes}(b) shows the distribution of the final number of votes received by front page stories that were submitted over a period of about two days in June 2006. The distribution is characteristic of `inequality of popularity', since a handful of stories become very popular, accumulating thousands of votes, while most others can only muster a few hundred votes. This distribution applies to front page stories only. Stories that are never promoted to the front page receive very few votes, in many cases just a single vote from the submitter. Such distributions are also called `long tailed' distributions. This means that in systems displaying such distributions extreme events, e.g., a story receiving many thousands of votes, occur much more frequently than would be expected if the underlying processes were Poisson or Gaussian in nature.

The long tail is  a  ubiquitous feature~\cite{Anderson06} of {human activity}. It is present in inequality of popularity of cultural artifacts, such as books and music albums~\cite{salganik06}, and also manifests itself in a variety of online behaviors, including tagging, where a few documents are tagged much more frequently than others,  collaborative editing on wikis~\cite{Kittur06}, and general social media usage~\cite{wilkinson08}. The same distribution of popularity was also observed in a sample of more than 30,000 stories promoted to Digg's front page over the course of a year~\cite{wu07}.

While unpredictability of popularity is more difficult to verify than in the controlled experiments of Salganik et al., it is reasonable to assume that a similar set of stories submitted to Digg on another day will end with radically different numbers of votes. In other words, while the distribution of the  final number of votes  these stories receive will look similar to the distribution in Figure~\ref{fig:votes}(b), the number of votes received by individual stories will be very different in the two realizations.

\subsection{Data collection}
We collected data for the study by scraping Digg's Web pages in May and June 2006. The May data set consists of stories that were submitted to Digg May 25-27, 2006. We followed these stories by periodically scraping Digg  to determine the number of votes stories received as a function of the time since their submission. We collected at least 4 such observations for each of 2152 stories, submitted by 1212 distinct users. Of these stories, 510, by 239 distinct users, were promoted to the front page. We followed the promoted stories over a period of several days, recording the number of votes the stories received. This May data set also records the location of the stories on the upcoming and front pages as a function of time.

The June data set consists of 201 stories promoted to the front page between June 27 and 30, 2006. For each story, we collected the names of its first 216 voters.

We focus our data collection on the early stages of story evolution -- from submission until shortly after promotion. The reason for this is that the Digg social network has a much larger effect on upcoming than front page stories due to the much more rapid addition of stories to the upcoming list. This large influx of stories makes it difficult for users to find a new story before it becomes hidden by the arrival of more stories. In this case, enhanced visibility via the network for fans of the submitter or early voters is particularly important, and a model of social dynamics has to account for it. In light of these observations, and for speeding up data collection, we focus on the early votes for stories.

Activity on Digg varies considerably over the course of a day, as seen in \fig{digg time}. Adjusting times by the cumulative activity on the site accounts for this variation and improves predictions~\cite{szabo09}. We define the ``Digg time'' between two events (e.g., votes on a story) as the total number of votes on front page stories during the time between those events. In our data set, there are on average about 2500 such votes per hour, with a range of about a factor of 4 in this rate during the course of a day. This behavior is similar to that seen in an extensive study of front page activity in 2007~\cite{szabo09}, and as in that study we scale the measure by defining a ``Digg hour'' to be the average number of front page votes in an hour, i.e., 2500 for our data set.
We evaluate the consequence of this variability by contrasting a model based on real time (in \sect{icwsm_model}) with one based on Digg time (in \sect{niche_model}).

\begin{figure}[t]
\centering
\includegraphics[width=\figwidth]{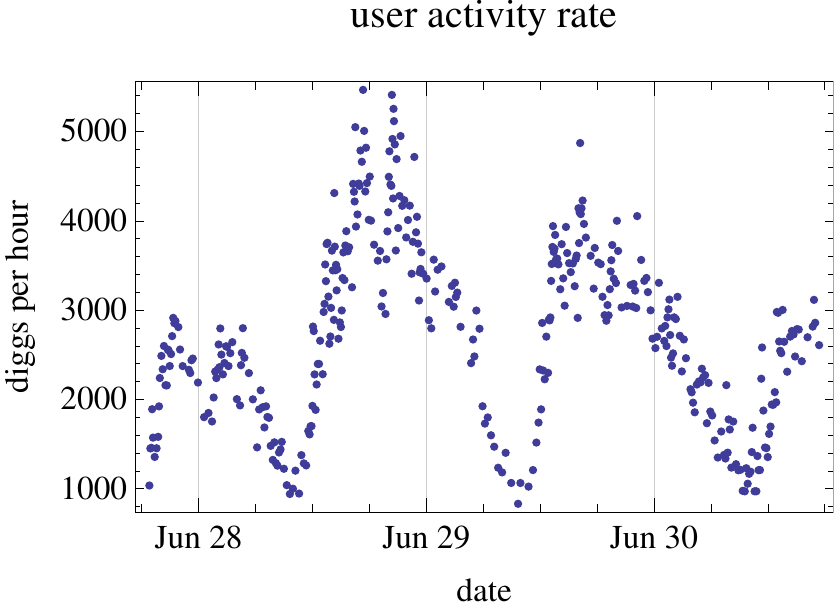}
\caption{Voting rate (diggs per hour) on front page stories at the end of June 2006. The indicated dates are the start of each day (0:00 GMT). The minimum in daily activity is around 9am GMT. Each point is the average voting rate for 100 successive votes.} \figlabel{digg time}
\end{figure}

In addition to voter activity, we also extracted a snapshot of the social network of the top-ranked 1020 Digg users as of June 2006. This data contained the names of each user's friends and fans. Since the original network did not contain information about all the voters in our data, we augmented it in February 2008 by extracting names of friends of about $15,000$ additional users.
Many of these users added friends between June 2006 and February 2008. Although Digg does not provide the time a new link was created, it lists the links in reverse chronological order and gives the date the friend joined Digg. By eliminating friends who joined Digg after June 30, 2006, we were able to reconstruct the fan links for all voters in our data. This data allows us to identify, for each vote, whether the user was a fan of any prior voter on that story, in which case the story would have appeared in the friends interface for that user.

Votes by fans account for 6\% of the votes in the June data set and about 3\% of the front page votes.

The data sets used in this and previous works were collected before Digg's API was introduced. Scraping Web pages to extract data had several issues. First, data had to be manually cleaned to ensure consistency. Second, since vote time stamps were not available on the Web page, we had to supplement June 2006 data by using the Digg API in October 2009 to obtain the time of each vote, the final number of votes the story received, and the time of promotion. In the intervening time, however, some of the users had deleted their accounts. Since we could not easily resolve the time of the vote of an inactive user, we had to delete these users from the voters list. We believe that the small fraction of data lost in this manner (less than 8\% of the data) does not adversely affect the modeling study.
However, in the future we plan to repeat the study on a much cleaner data set obtained through Digg API.

\section{Stochastic Models of Social Dynamics}
\label{sec:stochastic}
Rather than account for the inherent variability of individuals,
stochastic models focus on describing the macroscopic, or aggregate, behavior of the system, which can be described by {\em average} quantities. In the context of Digg, such quantities include average rate at which
users post new stories and vote on existing stories. Such
macroscopic descriptions often have a simple form and are
analytically tractable. Stochastic models do not reproduce the
results of a single observation --- rather, they describe the
`typical' behavior. These models are analogous to the approach used
in statistical physics, demographics and macroeconomics where the
focus is on relations among aggregate quantities, such as volume and
pressure of a gas, population of a country and immigration, or
interest rates and employment.

We represent each individual entity, whether a user or a story, as a stochastic process with a small number
of states. This abstraction captures much of the individual
complexity and environmental variability by casting user's decisions as inducing probabilistic
transitions between states. While this modeling framework applies to
stochastic processes of varying complexity, for simplicity, we focus
on simple processes that obey the Markov property, namely, a user
whose future state depends only on her present state and the input
she receives. A Markov process can be succinctly captured by a
\emph{state diagram} showing the possible states of the user and conditions for
transition between those states.

We assume that all users have the same set of states, and that transitions
between states depend only on the state and not the individual user. That is, the state captures the key relevant properties determining subsequent user actions. Then, the aggregate state of the system
can be described simply by the \emph{number} of individuals in
each state at a given time. That is, the system configuration at this time is
defined by the occupation vector: $ {\vec n} = (n_1, n_2,\ldots) $
where $n_k$ is the number of individuals in state $k$.
For example, in the context of a given story on Digg, one of the states for a user could be ``has voted for the story''. The component of the occupation vector corresponding to this state is the number of users who have voted for this story, without regard for which particular users those are.

The next step in developing the stochastic model is to summarize the
variation within the collection of histories with a probabilistic
description. That is, we characterize the possible occupation
vectors by the probability, $P({\vec n},t)$, the system is in
configuration ${\vec n}$ at time $t$. The evolution of $P({\vec
n},t)$, governed by the Stochastic Master
Equation~\cite{vankampen92}, is almost always too complex to be
analytically tractable. Fortunately we can simplify the problem by
working with the average occupation number, whose evolution is given
by the Rate Equation
\begin{equation}
\label{eqn-rate-3} \frac {d \langle n_k \rangle} {d t} = \sum_{j}
w_{jk}(\langle \vec n \rangle) \langle n_{j}\rangle - \langle n_k
\rangle \sum_{j}w_{kj}(\langle \vec n \rangle)
\end{equation}
where $\langle n_k \rangle$ denotes the average number of users in
state $k$ at time $t$, i.e., $\sum_{\vec n}n_k P({\vec n}, t)$ and
$w_{jk}(\langle \vec n \rangle)$ is the transition rate from
configuration $j$ to configuration $k$ when the occupation vector is
$\langle \vec n \rangle$.

Using the average of the occupation vector in the transition rates
is a common simplifying technique for stochastic models. A
sufficient condition for the accuracy of this approximation is that
variations around the average are relatively small. In many
stochastic models of systems with large numbers of components, variations are
indeed small due to many independent interactions among the
components. More elaborate versions of the stochastic approach give
improved approximations when variations are not small, particularly
due to correlated interactions~\cite{opper01}. User behavior on the
web, however, often involves distributions with long tails, whose typical
behaviors differ significantly from the average~\cite{barabasi05,wilkinson08}.
In this case we have no guarantee that the averaged approximation is
adequate. Instead we must test its accuracy for particular aggregate
behaviors by comparing model predictions with observations of actual
behavior, as we report below.

In the Rate Equation, occupation number $n_k$ increases due to
users' transitions from other states to state $k$, and decreases due
to transitions from the state $k$ to other states. The equations can
be easily written down from the user state diagram. Each state
corresponds to a dynamic variable in the mathematical model
--- the average number of users in that state --- and it is coupled
to other variables via transitions between states. Every transition
must be accounted for by a term in the equation, with transition
rates specified by the details of the interactions between users.

In summary, the stochastic modeling
framework requires specifying the aggregate states of interest for
describing the system and how individual user behaviors create
transitions among these states. The modeling approach is best suited
to cases where the users' decisions are mainly determined by a few
characteristics of the user and the information they have about the
system. These system states and transitions give the rate equations.
Solutions to these equations then give estimates of how aggregate
behavior varies in time and depends on the characteristics of the
users involved.

\section{A Model of Social Dynamics of Digg}
\sectlabel{icwsm_model}

Underlying a stochastic model of social dynamics is a behavioral model of an individual Web user. The behavioral model takes into account the choices a Web site's user interface allows users. Detailed data about human activity that can be collected from social media sites such as Digg allow us to parameterize the models and test them by comparing their predictions to the observed collective dynamics.

\begin{figure}[t]
  \begin{center}
  \includegraphics[height=1.9in]{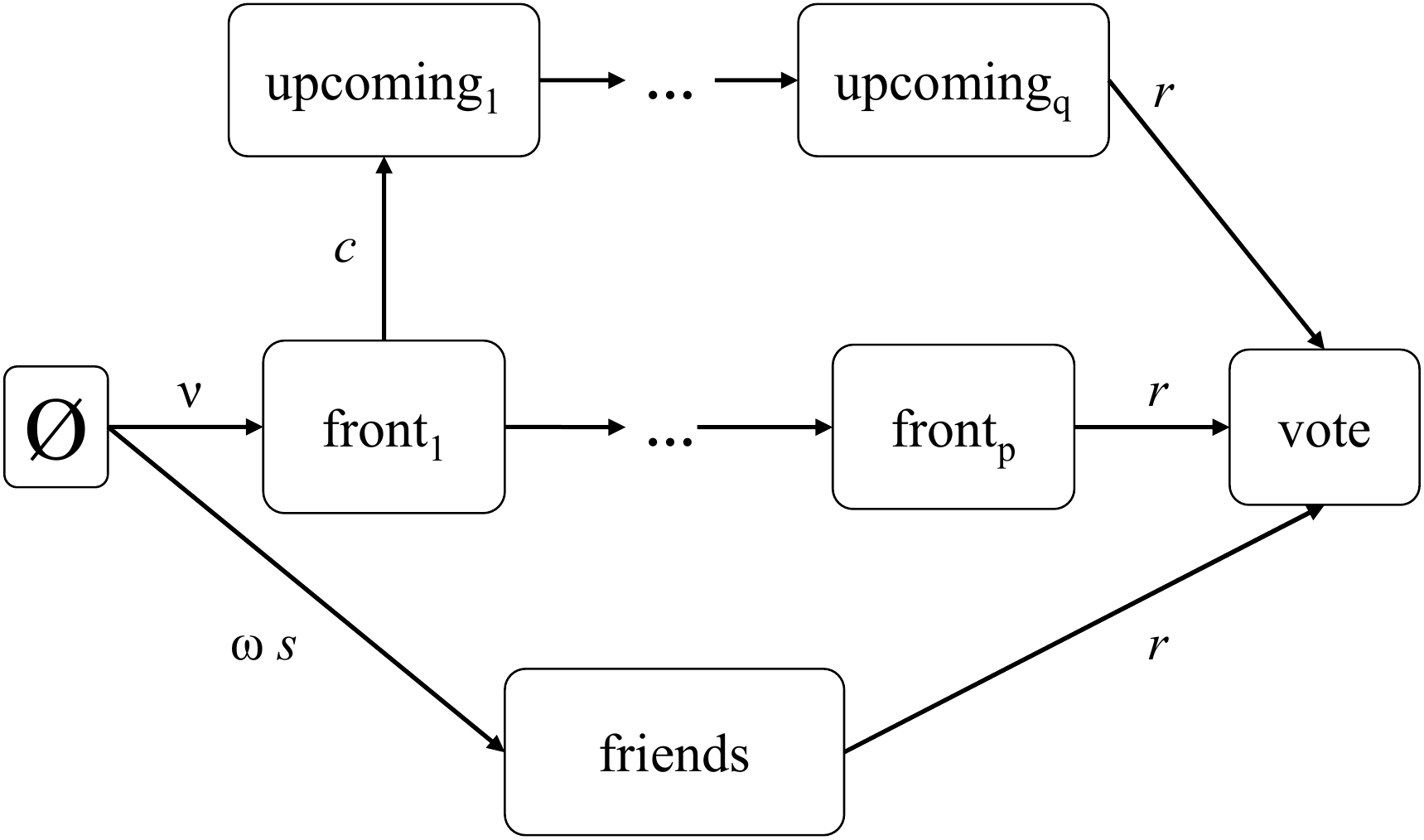}
  \end{center}
\caption{State diagram of user behavior for a single story. A user
starts in the $\emptyset$ state at the left, may find the story
through one of the three interfaces and may then vote on it. At a
given time, the story is located on a particular page of either the
upcoming or front page lists, not both. This diagram shows votes for
a story on either page $p$ of the front pages or page $q$ of the
upcoming pages. Only fans of previous voters can see the story
through the friends interface. Users in the friends, front or
upcoming states may choose to leave Digg, thereby returning to the
$\emptyset$ state (with those transitions not shown in the figure).
Users reaching the ``vote'' state remain there indefinitely and can
not vote on the story again. Parameters next to the arrows
characterize state transitions.} \label{fig:user-fsa}
\end{figure}

An earlier study of social dynamics of Digg~\cite{hogg09b} used a simple behavioral model that viewed each Digg user as a stochastic Markov process, whose state diagram with respect to a single story is shown in Fig~\ref{fig:user-fsa}. According to this model, a user visiting Digg can choose to browse the \emph{front} pages to see the recently promoted stories, \emph{upcoming} stories pages for the recently submitted stories, or use the \emph{friends} interface to see the stories her friends have recently submitted or voted for. She can select a story to read from one of these pages and, if she considers it interesting, \emph{vote} for it. The user's environment, the stories she is seeing, changes in time due to the actions of all the users.

We characterize the changing state of a story by three values: the
number of votes, $\voteTotal(t)$, the story has received by time $t$
after it was submitted to Digg, the list the story is in at time $t$
($upcoming$ or $front page$) and its location within that list, which
we denote by $q$ and $p$ for upcoming and front page lists,
respectively.

With Fig.~\ref{fig:user-fsa} as a modeling blueprint, we relate the
users' choices to the changes in the state of a single story. In
terms of the general rate equation (Eq.~\ref{eqn-rate-3}), the
occupancy vector ${\vec n}$ describing the aggregate user behavior
at a given time has the following components: the number of users
who see a story via one of the front pages, one of the upcoming
pages, through the friends pages, and number of users who vote for a
story, $\voteTotal$. Since we are interested in the number of users
who reach the vote state, we do not need a separate equation for
each state in Fig.~\ref{fig:user-fsa}: at a given time, a particular
story has a unique location on the upcoming or front page lists.
Thus, for simplicity, we can group the separate states for each list
in Fig.~\ref{fig:user-fsa}, and consider just the combined
transition for a user to reach the page containing the story at the
time she visits Digg. These combined transition rates depend on the
location of the story in the list, i.e., the value of $q$ or $p$ for
the story. With this grouping of user states, the rate equation for
$\voteTotal(t)$ is:
\begin{equation}\label{eq:diggs}
    \frac{d \voteTotal(t)}{d t} =r ( \frontRate(t) + \newRate(t) + \friendsRate(t) )
\end{equation}
\noindent where $r$ measures how interesting the story is, i.e., the
probability a user seeing the story will vote on it, and
$\frontRate$, $\newRate$ and $\friendsRate$ are the rates at which
users find the story via one of the front or upcoming pages, and
through the friends interface, respectively.

In this model, the transition rates appearing in the rate equation
depend on the time $t$ but not on the occupation vector.
Nevertheless, the model could be generalized to include such a
dependence if, for example, a user currently viewing an interesting
story not only votes on it but explicitly encourages people they
know to view the story as well.

\subsection{Story Visibility}

Before we can solve Eq.~\ref{eq:diggs}, we must model the rates at
which users find the story through the various Digg interfaces.
These rates depend on the story's location in the list. The
parameters of these models depend on user behaviors that are not
readily measurable. Instead, we estimate them using data collected
from Digg, as described below.

\paragraph{Visibility by position in list}
A story's visibility on the front page or upcoming stories lists
decreases as recently added stories push it further down the list.
The stories are shown in groups: the first page of each list
displays the 15 most recent stories, page 2 the next 15 stories, and
so on.

We lack data on how many Digg visitors proceed to page 2, 3 and so
on in each list. However, when presented with lists over multiple
pages on a web site, successively smaller fractions of users visit
later pages in the list. One model of users following links through
a web site considers users estimating the value of continuing at the
site, and leaving when that value becomes
negative~\cite{huberman98}. This model leads to an inverse Gaussian
distribution of the number of pages $m$ a user visits before leaving
the web site,
\begin{equation}\label{eq:stopping distribution}
e^{-\frac{\lambda  (m-\mu )^2}{2 m \mu ^2}} \sqrt{\frac{\lambda
   }{2 \pi m^3}}
\end{equation}
with mean $\mu$ and variance $\mu^3/\lambda$. This distribution
matches empirical observations in several web
settings~\cite{huberman98}. When the variance is small, for
intermediate values of $m$ this distribution approximately follows a
power law, with the fraction of users leaving after viewing $m$
pages decreasing as $m^{-3/2}$.

To model the visibility of a story on the $m^{th}$ front or upcoming
page, the relevant distribution is the fraction of users who visit
\emph{at least} $m$ pages, i.e., the upper cumulative distribution
of Eq.~\ref{eq:stopping distribution}. For $m>1$, this fraction is
\begin{equation}
\fractionToPage(m) = \frac{1}{2}\left( F_m(-\mu) - e^{2\lambda/\mu}
F_m(\mu) \right)
\end{equation}
where $F_m(x)=\erfc(\alpha_m (m-1+x)/\mu)$, $\erfc$ is the
complementary error function, and $\alpha_m =
\sqrt{\lambda/(2(m-1))}$. For $m=1$, $\fractionToPage(1)=1$.

The visibility of stories decreases in two distinct ways when a new
story arrives. First, a story moves down the list on its current
page. Second, a story at the $15^{th}$ position moves to the top of
the next page. For simplicity, we model these processes as
decreasing visibility, i.e., the value of $\fractionToPage(m)$,
through $m$ taking on fractional values within a page, i.e., $m=1.5$
denotes the position of a story half way down the list on the first
page. This model is likely to somewhat overestimate the loss of
visibility for stories among the first few  of the 15 items on a
given page since the top several stories are visible without
requiring the user to scroll down the page.

\begin{figure*}[t]
\begin{center}
  \begin{tabular}{cc}
  \includegraphics[height=1.8in]{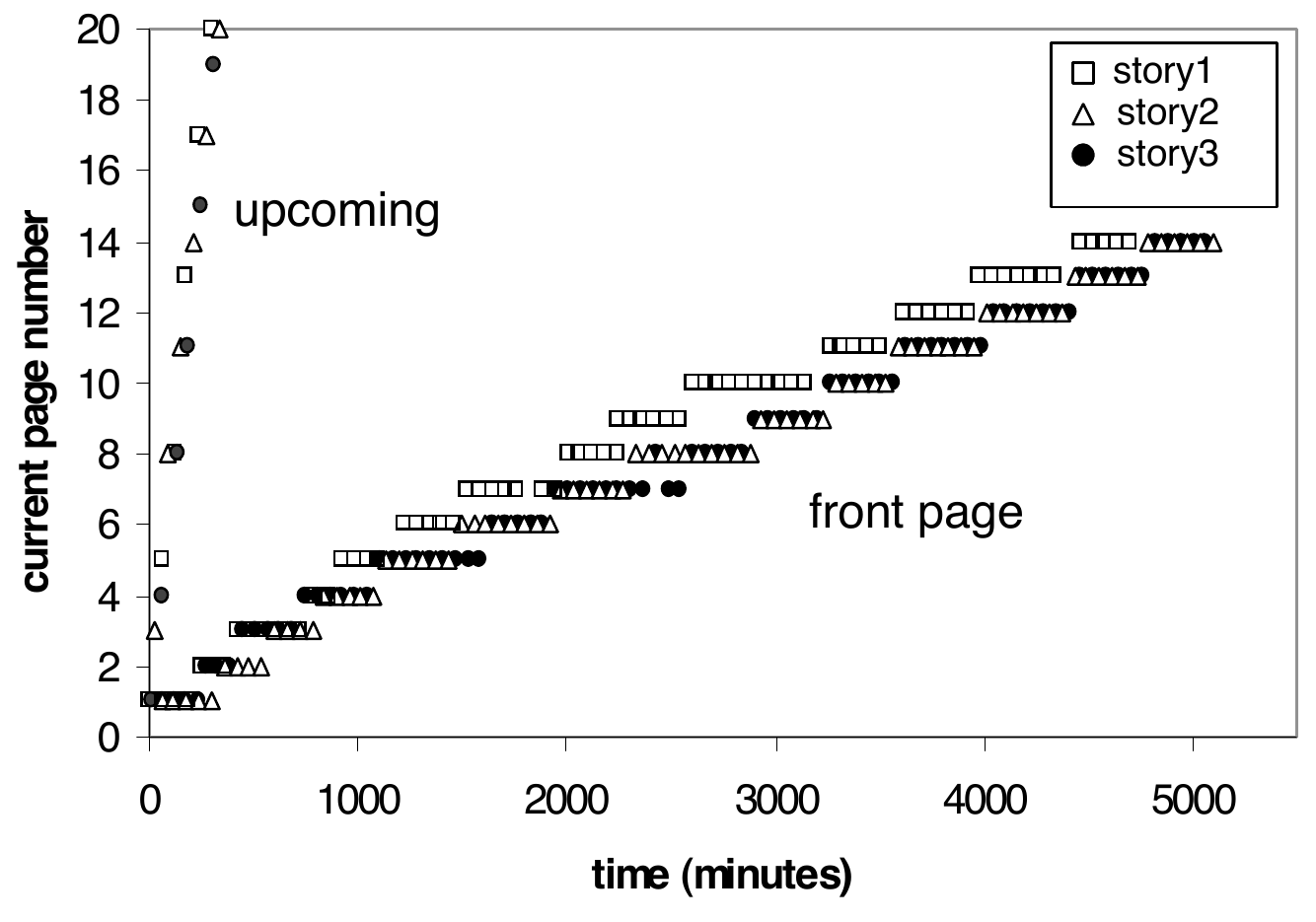} &
  \includegraphics[height=1.8in]{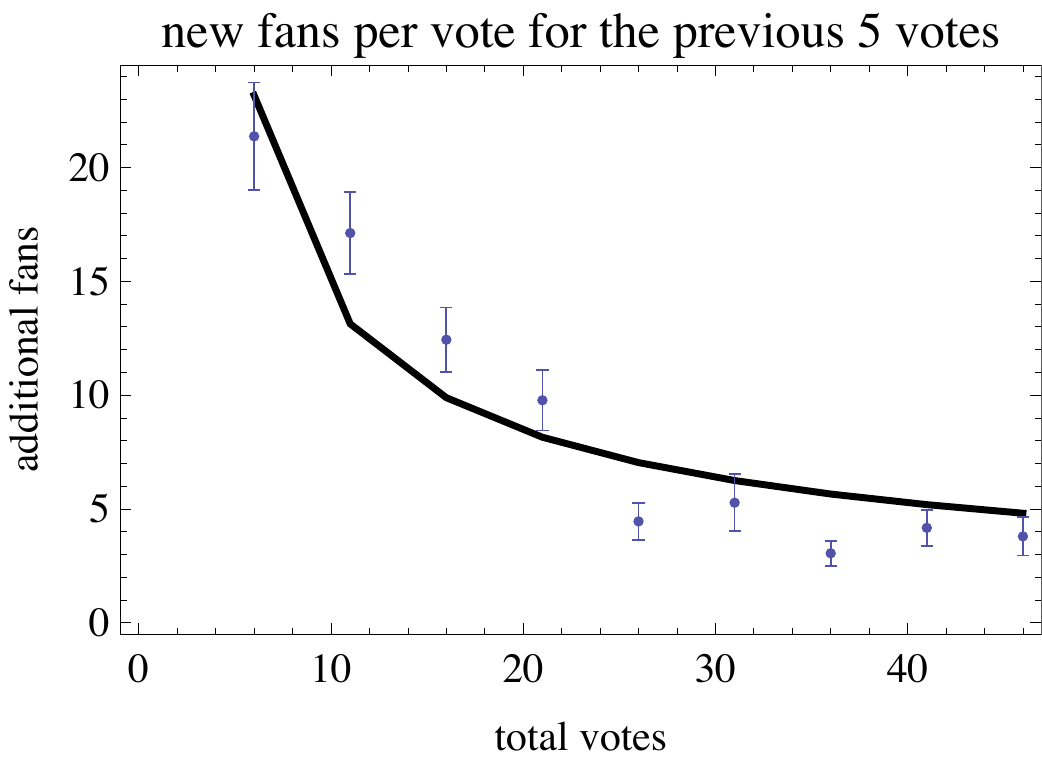} \\
  (a) & (b)
  \end{tabular}
\end{center}
\caption{(a) Current page number on the upcoming and front pages
vs.~time for three different stories. Time is measured from when the
story first appeared on each page, i.e., time it was submitted or
promoted, for the upcoming and front page points, respectively. (b)
Increase in the number of distinct users who can see the story
through the friends interface with each group of five new votes for
the first 46 users to vote on a story. The points are mean values
for 195 stories, including those shown in (a), and the curve is
based on Eq.~\ref{eq:Sv}. The error bars indicate the standard error
of the estimated means.} \label{fig:params}
\end{figure*}

\paragraph{List position of a story}
Fig.~\ref{fig:params}(a) shows how the page number of a story on the
two lists changes in time for three randomly chosen stories from our
data set. The behavior is close to linear when averaging over the daily activity variation (shown in \fig{digg time}). For simplicity in this model, we ignore this variation and take a story's page
number on the upcoming page $q$ and the front page $p$ at time $t$
to be~\cite{hogg09b}
\begin{eqnarray}
  p(t) &=& \frontPageGrowth (t-\Tpromotion)+1 \eqlabel{front page location} \\
  q(t) &=& \newPageGrowth t + 1 \eqlabel{upcoming page location}
\end{eqnarray}
where $\Tpromotion$ is the time the story is promoted to the front page (or $\infty$ if the story is never promoted)
and the slopes are given in
Table~\ref{parameters}. For a given story, $p(t)$ is only defined for times $t \geq \Tpromotion$ and $q(t)$ for $t < \Tpromotion$.
Since each page holds 15 stories, these rates are $1/15^{th}$ the
submission and promotion rates, respectively.

\paragraph{Front page and upcoming stories lists}
Digg prominently shows the stories on the front page. The upcoming
stories list is less popular than the front page. We model this fact
by assuming a fraction $c<1$ of Digg visitors proceed to the
upcoming stories pages.

We use a simple threshold to model how a story is promoted to the
front page. Initially the story is visible on the upcoming stories
pages. If and when the number of votes a story receives exceeds a
promotion threshold $h$, the story moves to the front page. This
threshold model approximates Digg's promotion algorithm as of May
2006, since in our data set we did not see any front page stories
with fewer than 44 votes, nor did we see any upcoming stories with
more than 42 votes. We take $h=40$ as an approximation to the
promotion algorithm.

\paragraph{Friends interface}
The friends interface allows the user to see the stories her friends
have (i) submitted, (ii) voted for, and (iii) commented on in the
preceding 48 hours. Although users can take advantage of all these
features, we only consider the first two. These uses of the friends
interface are similar to the functionality offered by other social
media sites: e.g., Flickr allows users to see the latest images his
friends uploaded, as well as the images a friend liked.

The fans of the story's submitter can find the story via the friends
interface. As additional people vote on the story, their fans can
also see the story. We model this with $s(t)$, the number of fans of
voters on the story by time $t$ who have not yet seen the story.
Although the number of fans is highly variable, the average number
of additional fans from an extra vote when the story has
$\voteTotal$ votes is approximately
\begin{equation}\label{eq:Sv}
\Delta s = a \voteTotal^{-b}
\end{equation}
where $a=51$ and $b=0.62$, as illustrated in
Fig.~\ref{fig:params}(b), showing the fit to the \emph{increment} in
average number of fans per vote over groups of 5 votes as given in
the data. Thus early voters on a story tend to have more new fans
(i.e., fans who are not also fans of earlier voters) than later
voters.

The model can incorporate any distribution for the times fans visit
Digg. We suppose these users visit Digg daily, and since they are
likely to be geographically distributed across all time zones, the
rate fans discover the story is distributed throughout the day. A
simple model of this behavior takes fans arriving at the friends
page independently at a rate $\omega$. As fans read the story, the
number of potential voters gets smaller, i.e., $s$ decreases at a
rate $\omega s$, corresponding to the rate fans find the story
through the friends interface, $\friendsRate$. We neglect additional
reduction in $s$ from fans finding the story without using the
friends interface.

Combining the growth in the number of available fans and its
decrease as fans return to Digg gives
\begin{equation}
\frac{d s}{d t} = -\omega s + a \voteTotal^{-b} \frac{d
\voteTotal}{d t}
\end{equation}
with initial value $s(0)$ equal to the number of fans of the story's
submitter, $S$.
This model of the friends interface treats the pool of fans
uniformly. That is we assume no difference in behavior, on average,
for fans of the story's submitter vs.~fans of other voters.

In summary, the rates in Eq.~\ref{eq:diggs} are\footnote{$\Theta(x)$ is a step function: $1$ when $x \ge 0$ and $0$ when $x<0$.}:
\begin{eqnarray*}
  \frontRate &=&  \visitRate \fractionToPage(p(t)) \, \Theta(\voteTotal(t)-h) \\
  \newRate &=& c \, \visitRate \fractionToPage(q(t)) \, \Theta(h-\voteTotal(t)) \Theta(24\hour-t)\\
  \friendsRate &=& \omega s(t)
\end{eqnarray*}
\noindent where $t$ is time since the story's submission and
$\visitRate$ is the rate users visit Digg. The first step function
in $\frontRate$ and $\newRate$ indicates that when a story has fewer
votes than required for promotion, it is visible in the upcoming
stories pages; and when $\voteTotal(t)>h$, the story is visible on
the front page. The second step function in $\newRate$ accounts for
a story staying in the upcoming list for at most $24$ hours. We
solve Eq.~\ref{eq:diggs} subject to initial condition
$\voteTotal(0)=1$, because a newly submitted story starts with a
single vote, from the submitter.

\subsection{Model Parameters}

\begin{table}[t]
\begin{center}
\begin{tabular}{l|l}
\hline parameter & value \\
\hline rate general users come to Digg & $\nu=600\,\mbox{users}/\hour$ \\ 
fraction viewing upcoming pages  & $c=0.3$ \\
rate a voters' fans come to Digg & $\omega=0.12/\hour$ \\ 
page view distribution  & $\mu=0.6$, $\lambda=0.6$ \\
fans per new vote & $a=51$, $b=0.62$ \\
vote promotion threshold    & $h=40$ \\
upcoming stories location  & $\newPageGrowth =3.60\,\mbox{pages}/\hour$ \\ %
front page location  & $\frontPageGrowth = 0.18\,\mbox{pages}/\hour$ \\ %
\hline \multicolumn{2}{c}{story specific parameters} \\
interestingness    & $r$ \\
number of submitter's fans  & $S$ \\
\end{tabular}
\end{center}
\caption{Model parameters.}\label{parameters}
\end{table}

The solutions of Eq.~\ref{eq:diggs} show how the number of votes
received by a story changes in time. The solutions depend on the
model parameters, of which only two parameters
--- the story's interestingness $r$ and number of fans the submitter
has $S$ --- change from one story to another. Therefore, we fix
values of the remaining parameters as given in
Table~\ref{parameters}.

As described above, we estimate some of these parameters (such as
the growth in list location, promotion threshold and fans per new
vote) directly from the data. The remaining parameters are not
directly given by our data set (e.g., how often users view the
upcoming pages) and instead we estimate them based on the model
predictions. The small number of stories in our data set, as well as
the approximations made in the model, do not give strong constraints
on these parameters. We selected one set of values giving a
reasonable match to our observations. For example, the rate fans
visit Digg and view stories via the friend's interface, given by
$\omega$ in Table~\ref{parameters}, has 90\% of the fans of a new
voter returning to Digg within the next 19 hours.
As another example of interpreting these parameter values, for the
page visit distribution the values of $\mu$ and $\lambda$ in
Table~\ref{parameters} correspond to about $1/6$ of the users
viewing more than just the first page.
These parameters could in principle be measured independently from
aggregate behavior with more detailed information on user behavior.
Measuring these values for users of Digg, or other similar web
sites, could improve the choice of model parameters.

\subsection{Results}

\begin{figure}[t]
\begin{center}
\includegraphics[width=3in]{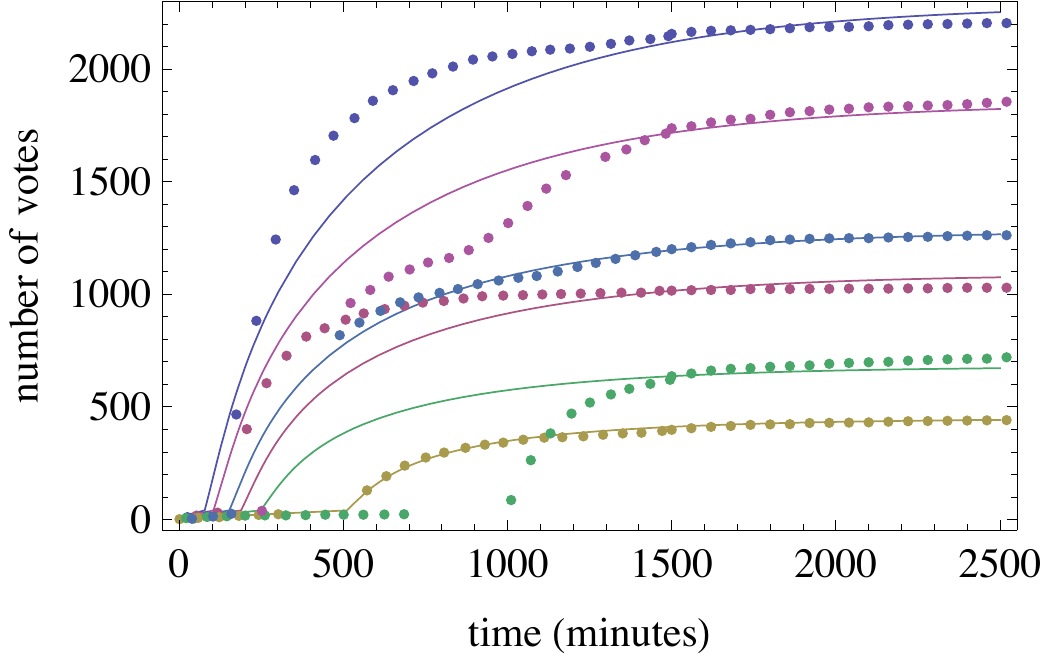}
\end{center}
\caption{Evolution of the number of votes received by six stories
compared with model solution. } \label{fig:predictions}
\end{figure}

\begin{table}[t]
\begin{center}
\begin{tabular}{ccc}
S & r & final votes \\
\hline 5 & 0.51 & 2229 \\
5 & 0.44 & 1921 \\
40 & 0.32 & 1297 \\
40 & 0.28 & 1039 \\
160 & 0.19 & 740 \\
100 & 0.13 & 458 \\
\end{tabular}
\end{center}
\caption{Parameters for the example stories, listed in decreasing order of
total votes received by the story and hence corresponding to the
curves in Fig.~\ref{fig:predictions} from top to
bottom.}\label{story parameters}
\end{table}

The model describes the behavior of all stories, whether or not they are promoted to the front page.
To illustrate the model results, we consider stories promoted to the front
page. Fig.~\ref{fig:predictions} shows the behavior of six
stories. For each story, $S$ is the number of fans of the story's
submitter, available from our data, and $r$ is estimated to minimize
the root-mean-square (\rms) difference between the observed votes
and the model predictions. Table~\ref{story parameters} lists these
values.

Overall there is qualitative agreement between the data and the model, indicating
that the features of the Digg user interface we considered can
explain the patterns of collective voting. Specifically, the model reproduces three generic behaviors of Digg stories: (1) slow initial growth in votes of upcoming stories; (2) more interesting stories (higher $r$) are promoted to the front page (inflection point in the curve)  faster and receive more votes than less interesting stories; (3)  however, as first described in \cite{Lerman07digg}, better connected users (high $S$) are more successful in getting their less interesting stories (lower $r$) promoted to the front page than poorly-connected users. These observations highlight a benefit
of the stochastic approach: identifying simple models of user
behavior that are sufficient to produce the aggregate properties of
interest.

The only significant difference between the data and the model is
visible in the lower two lines of Fig.~\ref{fig:predictions}. In the
data, a story posted by the user with $S=100$ is promoted before the
story posted by the user with $S=160$, but saturates at smaller
value of votes than the latter story. In the model, the story with
larger $r$ is promoted first and gets more votes.

Thus while the stochastic model is primarily intended to describe typical story behavior, we see it gives a reasonable match to the actual vote history of individual stories. Nevertheless, there are some cases where individual stories differ considerably from the model, particularly where an early voter happens to have an exceptionally large number of fans, thereby increasing the story's visibility to other users far more than the average number of new fans per vote. This variation, a consequence of the long-tail distributions involved in social media, is considerably larger than seen, for example, in most statistical physics applications of stochastic models. The effect of such large variations is an important issue for addressing the usefulness of the stochastic modeling approach for social media when applied to the behavior of individual stories.

\begin{figure}[t]
\begin{center}
\includegraphics[width=3in]{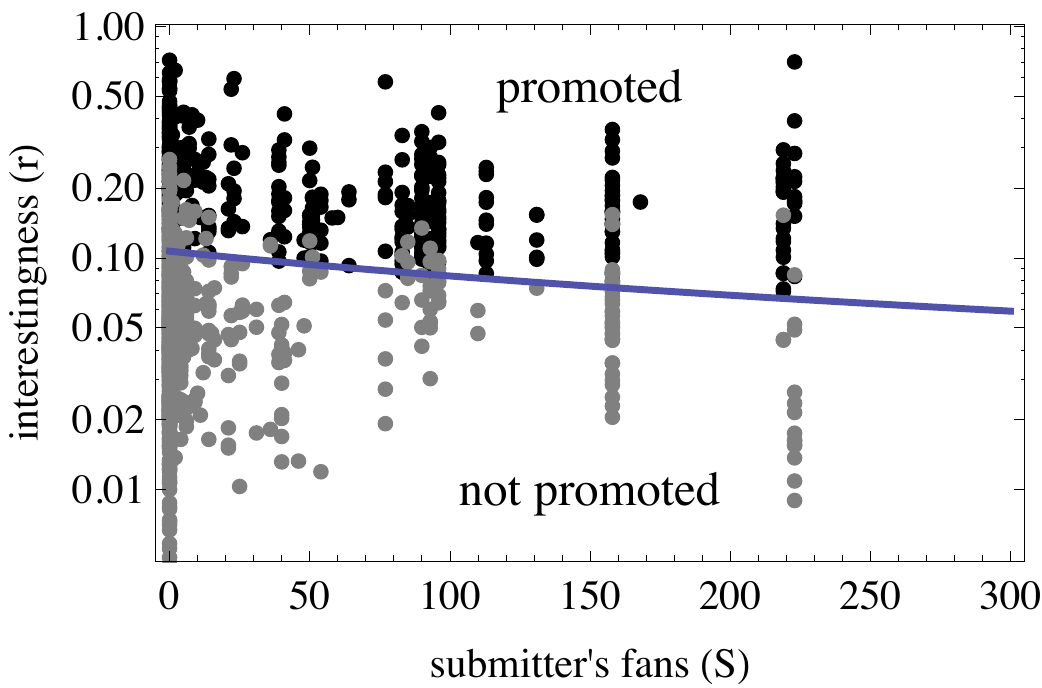}
\end{center}
\caption{Story promotion as a function of $S$ and $r$. The $r$
values are shown on a logarithmic scale. The model predicts stories
above the curve are promoted to the front page. The points show the
$S$ and $r$ values for the stories in our data set: black and gray
for stories promoted or not, respectively.}\label{fig:promotion}
\end{figure}

Fig.~\ref{fig:promotion} shows parameters required for a story to
reach the front page according to the model, and how that prediction
compares to the stories in our data set.
The model's prediction of whether a story is promoted is correct for
$95\%$ of the stories in our data set.
For promoted stories, the correlation between $S$ and $r$ is
$-0.13$, which is significantly different from zero ($p$-value less
than $10^{-4}$ by a randomization test).
Thus a story submitted by a poorly connected user (small $S$) tends
to need high interest (large $r$) to be promoted to the front
page~\cite{Lerman07digg}.

\begin{figure}[t]
\centering
 \includegraphics[height=1.8in]{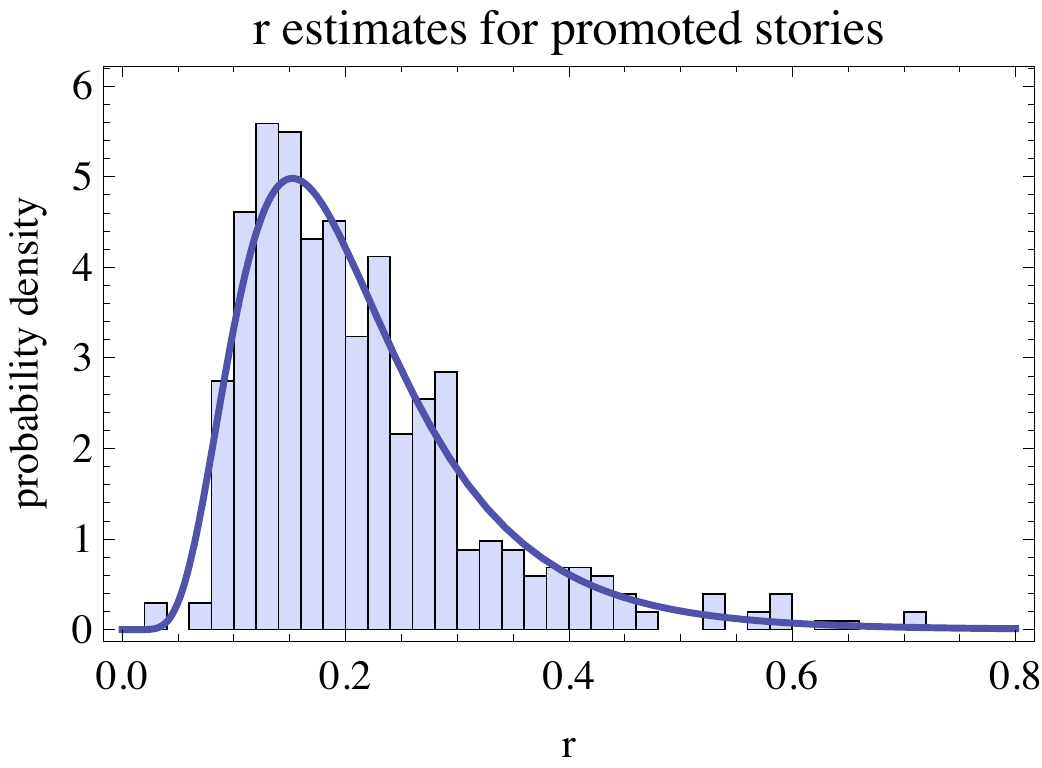}
 \caption{Distribution of interestingness (i.e., $r$ values) for the promoted
stories in our data set compared with the best fit lognormal
distribution.}
\label{fig:r-distribution}
\end{figure}

\begin{figure}[t]
\centering
 \includegraphics[width=\figwidth]{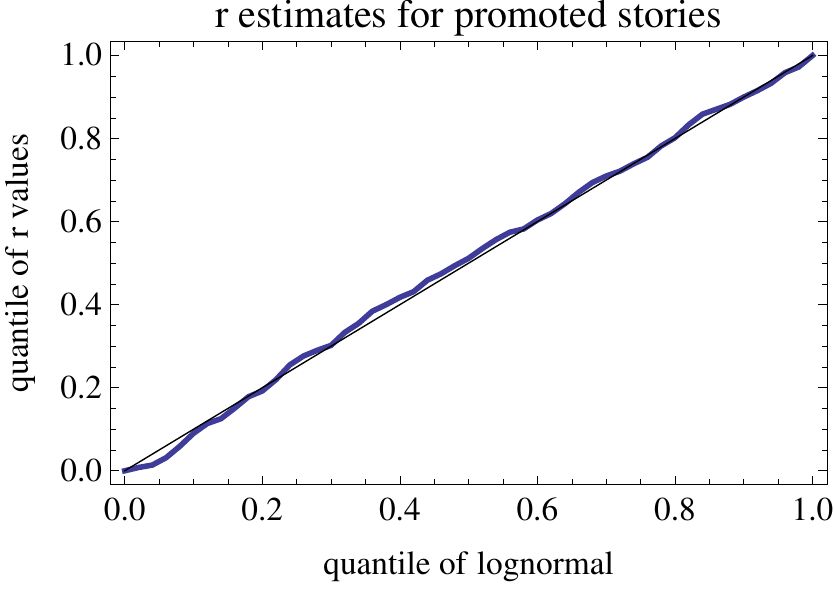}
 \caption{Quantile-quantile plot comparing observed
distribution of $r$ values with the lognormal distribution fit (thick curve). For comparison, the thin straight line from 0 to 1 corresponds to a perfect match between the data and the distribution.}
\label{fig:r-distribution quantiles}
\end{figure}

Figure~\ref{fig:r-distribution} shows the estimated $r$ values for the
510 promoted stories in our data set have a wide range of interestingness to users. That is, even after accounting for the variation in visibility of the stories, there remains a significant range in how well stories appeal to users. Specifically,
Fig.~\ref{fig:r-distribution quantiles} shows these $r$ values fit well
to a lognormal distribution
\begin{equation}\eqlabel{lognormal}
\Plognormal(\mu,\sigma;r) = \frac{1}{\sqrt{2\pi}\, r \sigma} \exp \left(  -\frac{(\mu-\log(r))^2}{2\sigma^2} \right)
\end{equation}
where parameters $\mu$ and $\sigma$ are the mean and standard deviation of $\log(r)$. For the distribution of interestingness values, the maximum likelihood estimates of the
mean and standard deviation of $\log(r)$ equal to $-1.67\pm0.04$ and
$0.47\pm0.03$, respectively, with the ranges giving the $95\%$
confidence intervals. A randomization test based on the
Kolmogorov-Smirnov statistic and accounting for the fact that the
distribution parameters are determined from the
data~\cite{clauset07} shows the $r$ values are consistent with this
distribution ($p$-value $0.35$). While broad distributions occur in
several web sites~\cite{wilkinson08}, our model allows factoring out
the effect of visibility due to the user interface from the overall
distribution of votes. Thus we can identify variation in users'
inclination to vote on a story they see.

The simple model described in this section gives a reasonable qualitative account of how user
behavior leads to stories' promotion to the front page and the
eventual saturation in the number of votes they receive due to their
decreasing visibility. In the section below we show how additional properties of the interface and
user population can be added to the model for a more accurate
analysis of the aggregate behavior. For example, submitter's fans may find
the story more interesting than the general Digg audience,
corresponding to different $r$ values for these groups of users.
In addition, we modeled users coming to Digg independently with uniform rates $\nu$
and $\omega$. In fact, the rates vary systematically over hours and
days~\cite{szabo09} as shown in  \fig{digg time}, and individual users have a wide range in time
between visits~\cite{vazquez06}. In our model, this variation gives
time-dependent values for $\nu$, describing the rate users come to
Digg, and $\frontPageGrowth$ and $\newPageGrowth$, which relate to
the rate new stories are posted and promoted.

The ability of the stochastic approach to incorporate additional details in the user
models illustrates its value in providing insights
into how aggregate behavior arises from the users, in contrast to
models that evaluate regularities in the aggregate behaviors~\cite{wu07}. In
particular, user models can help distinguish aggregate behaviors
arising from intrinsic properties of the stories (e.g., their
interestingness to the user population) from behavior due to the
information the web sites provides, such as ratings of other users
and how stories are placed in the site, i.e., visibility. Finally, stochastic models have not only explanatory, but also predictive power.

\section{A Model of Social Voting with Niche Interests}
\sectlabel{niche_model}

\begin{figure}[tb]
\centering
\includegraphics[viewport=75 125 750 540,clip,width=\figwidth]{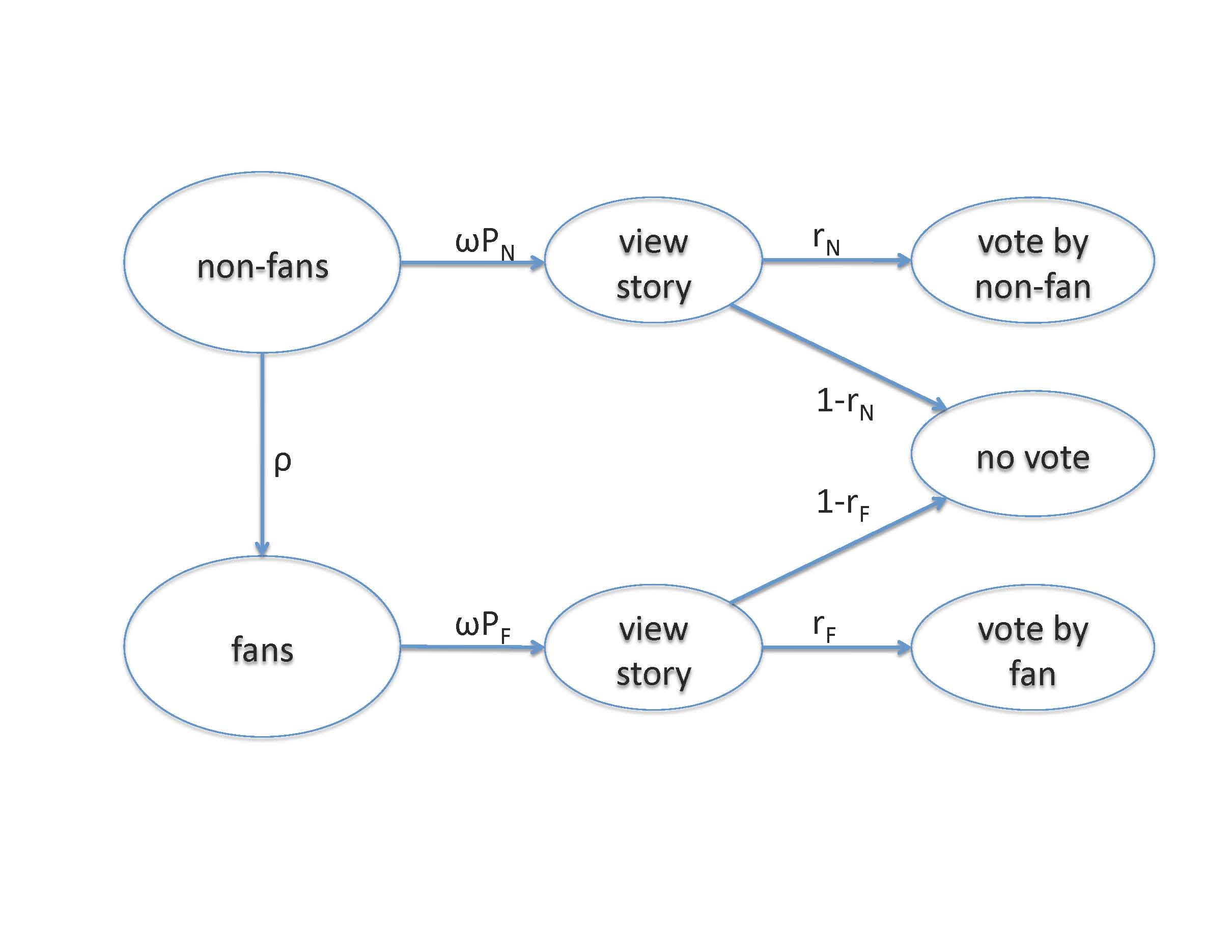}
\caption{State diagram for a user. The submitter provides a story's first vote.
The initial set of fans consists of the submitter's fans; other users are initially non-fans. Fans and non-fans have different probabilities to see and vote on the story.
With each vote, a non-fan user who is a fan of that voter moves into the \emph{fans} state.
This state transition is caused by the votes of other users: a user moving from the \emph{non-fans} to \emph{fans} state is not aware of that change until later visiting Digg and seeing the story in the friends interface.}\figlabel{state diagram}
\end{figure}

\begin{figure}[tb]
\centering
\includegraphics[viewport=60 380 600 530,clip,width=\figwidth]{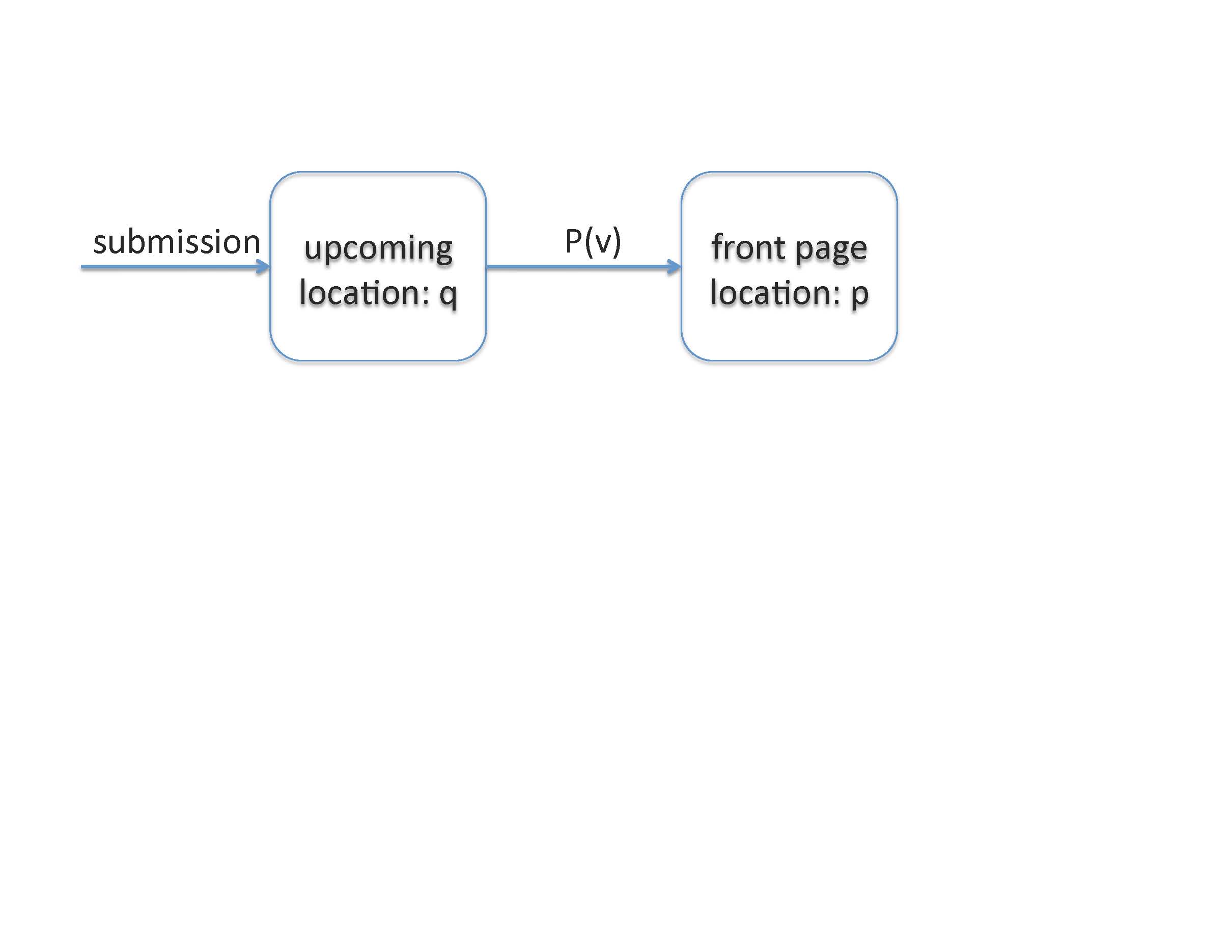}
\caption{State diagram for a story. A story starts at the top of the upcoming pages, with location $q=1$. The location increases with each new submission. An upcoming story with $v$ votes is promoted with probability $P(v)$. A promoted story starts at the top of the front pages, with location $p=1$. The location increases as more stories are promoted. A story not promoted within a day is removed (not shown).
}\figlabel{state diagram story}
\end{figure}

To investigate differences among voters with respect to the friends network, we extend the previous stochastic model to distinguish votes from fans and non-fans. The model considers the joint behavior of users and the location of the story on the web site. \fig{state diagram} shows the user states and the stochastic transitions between them. Stories are on either the upcoming or front pages, as shown in \fig{state diagram story}.
This leads to a description of the average rates of growth for votes from fans and non-fans of prior voters, $\fanVotes$ and $\nonfanVotes$, respectively:
\begin{eqnarray}\eqlabel{fan votes}
\frac{d \fanVotes}{dt} &=& \omega \Rfan \Pfan \fans \\
\eqlabel{nonfan votes}
\frac{d \nonfanVotes}{dt} &=& \omega \Rnonfan \Pnonfan \nonfans
\end{eqnarray}
where $t$ is the Digg time since the story's submission and $\omega$ is the average rate a user visits Digg (measured as a rate per unit Digg time). $\nonfanVotes$ includes the story's submitter.  $\Pfan$ and $\Pnonfan$ denote the story's \emph{visibility} and $\Rfan$ and $\Rnonfan$ denote the story's \emph{interestingness} to users who are fans or not of prior voters, respectively. Visibility depends on the story's state (e.g., whether it has been promoted), as discussed below. Interestingness is the probability a user who sees the story will vote on it. Nominally people become fans of those whose contributions they consider interesting, suggesting fans likely have a systematically higher interest in stories. Our model accounts for this possibility with separate interestingness values for fans and non-fans.

In contrast to the model of \sect{icwsm_model} where time $t$ denoted real time since story submission, we now use $t$ to denote the ``Digg time'' since submission, thereby accounting for the daily variation in activity. Using Digg time reduces the variation in the rate users visit Digg, thereby improving the match to the assumed constant rate $\omega$ used in the model. Moreover, a detailed examination of the page locations of the stories in our data set, shows systematic variation in the time stories spend on each page corresponding to the daily activity variation used to define Digg time. Thus using Digg time improves the accuracy of the linear growth in location given in \eq{front page location} and \eqbare{upcoming page location}.

These voting rates depend on $\fans$ ($\nonfans$), the numbers of users who have not yet seen the story and who are (are not) fans of prior voters. The quantities change as users see and vote on the story according to
\begin{eqnarray}\eqlabel{fans}
\frac{d \fans}{dt} &=& -\omega \Pfan \fans + \probUserIsAFan \nonfans \frac{d \Votes}{dt}\\
\eqlabel{nonfans}
\frac{d \nonfans}{dt} &=& -\omega \Pnonfan \nonfans - \probUserIsAFan \nonfans \frac{d \Votes}{dt}
\end{eqnarray}
with $\Votes = \fanVotes + \nonfanVotes$ the total number of votes the story has received.
The quantity $\probUserIsAFan$ is the probability a user who has not yet seen the story and is not a fan of a prior voter is a fan of the most recent voter. For simplicity, we treat this probability as a constant over the voters, thus averaging over the variation due to clustering in the social network and the number of fans a user has.
The first term in each of these equations is the rate the users see the story. The second terms arise from the rate the story becomes visible in the friends interface of users who are not fans of previous voters but are fans of the most recent voter.

Initially, the story has one vote (from the submitter) and the submitter has $S$ fans, so $\fanVotes(0)=0$, $\nonfanVotes(0)=1$, $\fans=S$ and $\nonfans=\Users-S-1$ where $\Users$ is the total number of active users at the time the story is submitted. Over time, a story becomes less visible to users as it moves down the upcoming or (if promoted) front page lists, thereby attracting fewer votes and hence fewer new fans of prior voters.

We use the same visiting rate parameter, $\omega$, for users who are and are not fans of prior voters since there is only a small correlation between voting activity and the number of fans across all the stories in our data set, as illustrated in \fig{votes vs fans}. Moreover,  many highly active users do not participate in the social network at all (i.e., have neither fans nor friends). Among all users, the correlation between number of votes and number fans is $0.15$. More specifically, we assume that with respect to votes on a single story, fans of those voters aren't systematically more likely to visit Digg than other users, such as fans of voters on other stories or users without fans or friends.

\begin{figure}[tb]
\centering
\includegraphics[width=\figwidth]{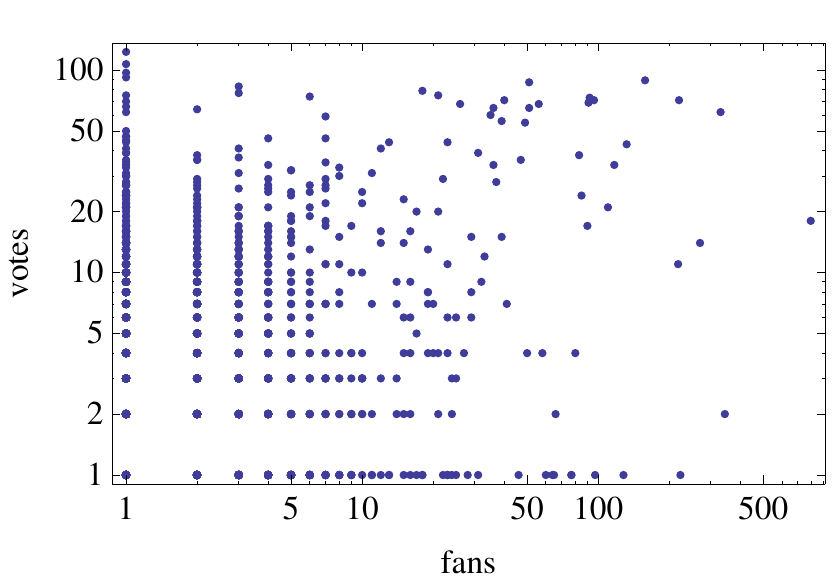}
\caption{Comparison of activity (number of votes) and number of fans for each of the 3436 users with at least one vote and one fan.
}\figlabel{votes vs fans}
\end{figure}

\subsection{Story Visibility}

We assume a fan easily sees the story via the friends interface, so $\Pfan=1$, as in the previous model~\cite{hogg09b}.
Users who are not fans of prior voters must find the story on the front or upcoming pages. Thus $\Pnonfan$ depends on how users navigate through these pages and the story's location at the time the user visits Digg. As with the previous model, we use Eq.~\ref{eq:stopping distribution} to describe this behavior.
\paragraph{List position of a story}
The page number of a story on the upcoming page $q$ and the front page $p$ at time $t$ is given by \eq{front page location} and \eqbare{upcoming page location}, with $t$ now interpreted as Digg time. The slopes, given in
\tbl{parameters}, are the same as with the previous model which averaged over the daily variation in activity. Since each page holds 15 stories, these
rates are $1/15^{th}$ the story submission and promotion rates,
respectively.

Since upcoming stories are less popular than the front page, our model has a fraction $c<1$ of Digg visitors viewing the upcoming stories pages.
Combining these effects, we take the visibility of a story at position $p$ in the front page list to be $\Pnonfan=\fractionToPage(p)$, whereas a story at position $q$ in the upcoming page list is $c \fractionToPage(q)$~\cite{hogg09b}.

\paragraph{Promotion to the front page}
Promotion to the front page appears to depend mainly on the number of votes the story receives.
We model this process by the probability $P(v)$ an upcoming story is promoted after its $v^{th}$ vote. We take $P(1)=0$, i.e., a story is not promoted just based on the submitter's vote. The probability a story is not promoted by the time it receives $v$ votes is $\prod_{i=1}^v (1-P(i))$. Stories not promoted are eventually removed, typically 24 hours after submission.

Based on our data, \fig{promotion} shows the probability $P(v)$ an upcoming story is promoted after $v$ votes conditioned on it not having been promoted earlier. We find a significant spread in the number of votes a story has when it is promoted. For predicting whether and when a story will be promoted in our model, we use a logistic regression fit to these values, as shown in the figure. This contrasts with the step function for promotion at 40 votes used in the previous model~\cite{hogg09b}.

\begin{figure}[tb]
\centering
\includegraphics[width=\figwidth]{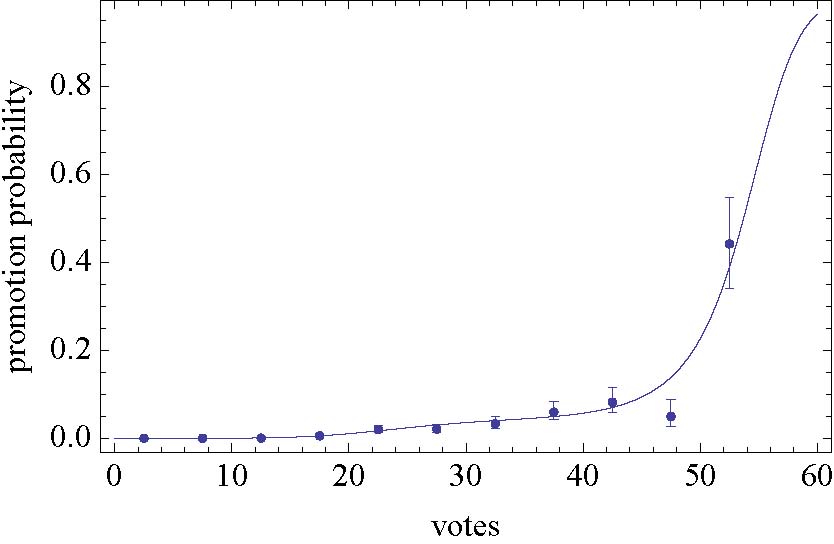}
\caption{Probability for promotion before the next vote for an upcoming story as a function of the number of votes. The error bars indicate the 95\% confidence intervals for the estimates. The curve is a logistic fit.}\figlabel{promotion}
\end{figure}

\paragraph{Friends interface}

The fans of the story's submitter can find the story via the friends
interface. As additional people vote on the story, their fans can
also see the story. We model this with $\fans(t)$, the number of fans of
voters on the story by time $t$ who have not yet seen the story.
Although the number of fans is highly variable, we use the average number
of additional fans from an extra vote, $\probUserIsAFan \nonfans$, in \eq{fans}.

\subsection{Parameter Estimation}

Since we observe votes, not visits to Digg, there is some ambiguity in the rate $\omega$ and the interestingness values $\Rfan$, $\Rnonfan$.
For example, a given value of $\omega \Rfan$ could arise from users often visiting Digg but rarely voting on stories, or less frequent visits with a high chance of voting during each visit. This arbitrary scaling does not affect our focus on the relative behavior of fans and non-fans. For definiteness, we pick a specific value for $\omega$ and give interestingness values relative to that choice.

We used the May data to estimate the story location parameters $\newPageGrowth$ and $\frontPageGrowth$. Their values correspond to 54 and 2.7 stories per hour submitted and promoted, respectively. 

\subsubsection{Estimating parameters from observed votes}
In our model, story location affects visibility only for non-fan voters since fans of prior voters see the story via the friends interface. Thus we use just the non-fan votes to estimate visibility parameters, via maximum likelihood. Specifically, we use the non-fan votes for 16 stories in the June data set to estimate $c$ and the ``law of surfing'' parameters $\mu$ and $\lambda$. We then use fan votes for these stories to evaluate the  probability a user is a fan of a new voter, $\probUserIsAFan$. Separating votes by the different interfaces by which users find stories provides more precise estimation than the prior model~\cite{hogg09b}.

This estimation involves comparing the observed votes to the voting rate from the model. As described above, the model uses rate equations to determine the average behavior of the number of votes. A simple approach to relate this average to the observed number of votes is to assume the votes from non-fan users form a Poisson process whose expected value is $d \nonfanVotes(t)/dt$, given by  \eq{nonfan votes}. This rate changes with time and depends on the model parameters.

For a Poisson process with a constant rate $v$, the probability to observe $n$ events in time $T$ is the Poisson distribution $e^{-v T} (v T)^n/n!$. This probability depends only on the \emph{number} of events, not the specific times at which they occur. Thus estimating a constant rate involves maximizing this expression, which gives $v = n/T$, i.e., the maximum-likelihood estimate of the rate for a constant Poisson process is equal to the average rate of the observed events.

In our case, the voting rate changes with time, requiring a generalization of this estimation. Specifically consider a Poisson process with nonnegative rate $v(t)$ which depends on one or more parameters to be estimated. Thus in a small time interval $(t,t+\Delta T)$, the probability for a vote is $v(t) \Delta t$, and this is independent of votes in other time intervals, by the definition of a Poisson process. Suppose we observe $n$ votes at times $0< t_1 < t_2, \ldots < t_n < T$ during an observation time interval $(0,T)$. Considering small time intervals $\Delta t$ around each observation, the probability of this observation is
\begin{eqnarray*}
P(\mbox{no vote in $(0,t_1)$}) v(t_1)\Delta t  & \times \\
P(\mbox{no vote in $(t_1,t_2)$}) v(t_2)\Delta t  &\times \\
 \ldots & \\
 P(\mbox{no vote in $(t_{n-1},t_n)$}) v(t_n)\Delta t  &\times \\
  P(\mbox{no vote in $(t_n,T)$})
\end{eqnarray*}
The probability for no vote in the interval $(a,b)$ is
\begin{displaymath}
\exp \left( -\int_a^b v(t) dt \right)
\end{displaymath}
Thus the log-likelihood for the observed sequence of votes is
\begin{displaymath}
 -\int_0^T v(t) dt  + \sum_i \log v(t_i)
\end{displaymath}
The maximum-likelihood estimation for parameters determining the rate $v(t)$ is a trade-off between these two terms: attempting to minimize $v(t)$ over the range $(0,T)$ to increase the first term while maximizing the values $v(t_i)$ at the specific times of the observed votes. If $v(t)$ is constant, this likelihood expression simplifies to $-v T + n \log v$ with maximum at $v=n/T$ as discussed above for the constant Poisson process. When $v(t)$ varies with time, the maximization selects parameters giving relatively larger $v(t)$ values where the observed votes are clustered in time.

In our case, we combine this log-likelihood expression from the votes on several stories, and maximize the combined expression with respect to the story-independent parameters of the model, with the interestingness parameters determined separately for each story.

\subsubsection{Estimating number of active users}
Our model involves a population of ``active users'' who visit Digg during our sample period.
Specifically, the model uses the rate users visit Digg, $\omega \Users$. We do not observe visits in our data, but can infer the relevant number of active users, $\Users$, from the heterogeneity in the number of votes by users. The June data set consists of 16283 users who voted at least once during the sample period. \fig{user activity} shows the distribution of this activity on front page stories. Most users have little activity during the sample period, suggesting a large fraction of users vote infrequently enough to never have voted during the time of our data sample. This behavior can be characterized by an activity rate for each user. A user with activity rate $\nu$ will, on average, vote on $\nu T$ stories during a sample time $T$.  We model the observed votes as arising from a Poisson process whose expected value is $\nu T$ and the heterogeneity arising from a lognormal distribution of user activity rates~\cite{hogg09c}. This model gives rise to the extended activity distribution while accounting for the discrete nature of the observations. The latter is important for the majority of users who have low activity rates so will vote only a few times, or not at all, during our sample period.

Specifically, for $n_k$ users with $k$ votes during the sample period, this mixture of lognormal and Poisson distributions~\cite{bulmer74, miller07} gives the log-likelihood of the observations as
\begin{displaymath}
\sum_k n_k \log P(\mu,\sigma;k)
\end{displaymath}
where $P(\mu,\sigma;k)$ is the probability of a Poisson distribution to give $k$ votes when its mean is chosen from a lognormal distribution $\Plognormal$ with parameters $\mu$ and $\sigma$. From \eq{lognormal},
\begin{displaymath}
P(\mu,\sigma;k) = \frac{1}{{\sqrt{2 \pi } \sigma  k!}} \int_0^\infty   \rho ^{k-1} e^{-\frac{(\log (\rho )-\mu )^2}{2 \sigma ^2}-\rho
   } d\rho
\end{displaymath}
for integer $k\geq 0$. We evaluate this integral numerically. In terms of our model parameters, the value of $\mu$ in this distribution equals $\nu T$.

Since we don't observe the number of users who did not vote during our sample period, i.e., the value of $n_0$, we cannot maximize this log-likelihood expression directly. Instead, we use a zero-truncated maximum likelihood estimate~\cite{hilbe08} to determine the parameters $\mu$ and $\sigma$ for the vote distribution of \fig{user activity}. Specifically, the fit is to the probability of observing $k$ votes conditioned on observing at least one vote. This conditional distribution is $P(\mu,\sigma;k)/(1-P(\mu,\sigma;0))$ for $k>0$, and the corresponding log-likelihood is
\begin{displaymath}
\sum_{k>0} n_k \log P(\mu,\sigma;k) - U_{+} \log(1-P(\mu,\sigma;0))
\end{displaymath}
where $U_{+}$ is the number of users with at least one vote in our sample, i.e., 16283.
Maximizing this expression with respect to the distribution's parameters $\mu$ and $\sigma$ gives
$\nu T$ lognormally distributed with the mean and standard deviation of $\log(\nu T)$ equal to $-2.06\pm0.03$ and $1.82\pm0.03$, respectively. With these parameters, $P(\mu,\sigma; 0)=0.757$, indicating about 3/4 of the users had sufficiently low, but nonzero, activity rate that they did not vote during the sample period. We use this value to estimate $\Users$, the number of active users during our sample period: $U = U_{+}/(1-P(\mu,\sigma; 0))$.

Based on this fit, the curve in \fig{user activity} shows the  expected number of users with each number of votes, i.e., the value of $\Users P(\mu,\sigma;k)$ for $k>0$. This is a discrete distribution: the lines between the expected values serve only to distinguish the model fit from the points showing the observed values.
A bootstrap test~\cite{efron79} based on the
Kolmogorov-Smirnov (KS) statistic shows the vote counts are consistent with this
distribution ($p$-value $0.48$). This test and the others reported in this paper account for the fact that we fit the distribution parameters to the
data~\cite{clauset07}.

\begin{figure}[tb]
\centering
\includegraphics[width=\figwidth]{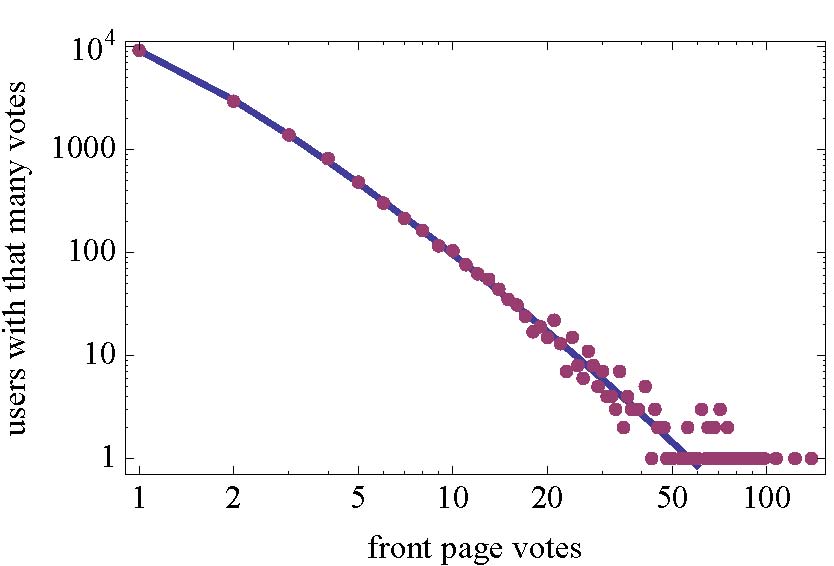}
\caption{User activity distribution on logarithmic scales. The curve shows the fit to the model described in the text.}\figlabel{user activity}
\end{figure}

\begin{table}[tb]
\begin{center}
{\small 
\begin{tabular}{l|l}
\hline parameter & value \\
\hline average rate each user visits Digg & $\omega=0.2\,/\hour$ \\
number of active users & $\Users = 70,000$ \\
fraction viewing upcoming pages  & $c=0.065$ \\
page view distribution  & $\mu=6.3$ \\
				& $\lambda=0.14$ \\
probability a user is a voter's fan & $\probUserIsAFan=9.48 \times 10^{-6}$ \\
upcoming stories location  & $\newPageGrowth =3.60\,\mbox{pages}/\hour$ \\ %
front page location  & $\frontPageGrowth = 0.18\,\mbox{pages}/\hour$ \\ %
\hline \multicolumn{2}{c}{story specific parameters\rule{0pt}{10pt}} \\
interestingness to fans    & $r_F$ \\
interestingness to non-fans    & $r_N$ \\
number of submitter's fans  & $S$ \\
\end{tabular}
}
\end{center}
\caption{Model parameters, with times in ``Digg hours''.}\tbllabel{parameters}
\end{table}

\subsubsection{Estimated parameters}
\tbl{parameters} lists the estimated parameters.
We estimate $\Rfan$ and $\Rnonfan$ for each story from its fan and non-fan votes.

The page view distribution seen in this data set
indicates users who choose to visit the upcoming pages tend to explore those pages fairly deeply. This contrasts with the more limited exploration, i.e., smaller value of $\mu$, seen in the May data set which included votes well after promotion~\cite{hogg09b}. This suggests differing levels of perseverance of users who visit the upcoming stories compared to the majority of users who focus on front page stories. Alternatively, there could be other ways non-fan users find content that has already moved far down the list of stories.

\subsection{Results}

\begin{figure}[tb]
\centering
\includegraphics[width=\figwidth]{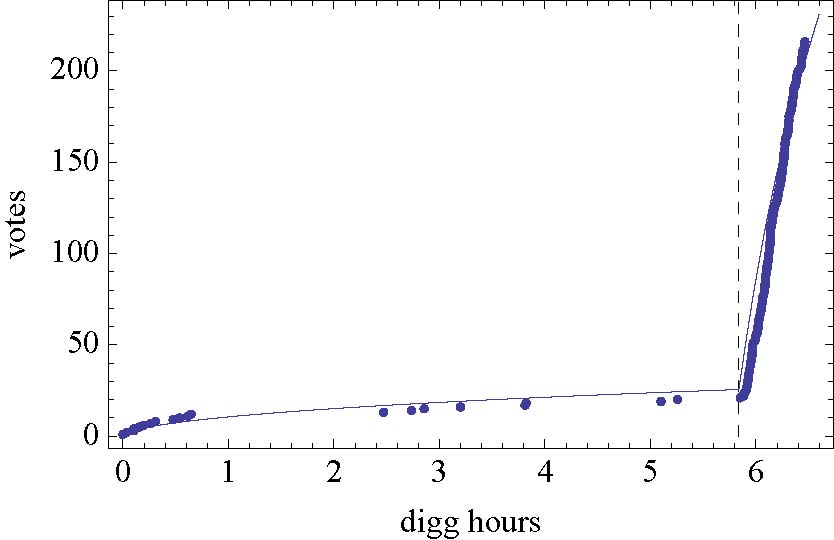}
\caption{Voting behavior: the number of votes vs.~time, measured in Digg hours, for a promoted story in June 2006. The curve shows the corresponding solution from our model and the dashed vertical line indicates when the story was promoted to the front page. This story eventually received 2566 votes.}\label{vote example}
\end{figure}

Figure~\ref{vote example} compares the solution of the rate equations with the actual votes for one story. The model correctly reproduces the dynamics of voting while the story is on the upcoming stories list and immediately after promotion.


We use the model to evaluate systematic differences in story interestingness between fans and non-fans, with the resulting distribution of values shown in \fig{r values}. The interestingness values for fans and non-fans of prior voters each have a wide range of values, but the interestingness to fans is generally much higher than to non-fans. Both sets of values fit well to lognormal distributions, as indicated in \fig{r value quantiles}.
Specifically, the $\Rnonfan$ values fit well to
a lognormal distribution with maximum likelihood estimates of the
mean and standard deviation of $\log(\Rnonfan)$ equal to $-4.0\pm0.1$ and
$0.63\pm0.07$, respectively, with the ranges giving the $95\%$
confidence intervals. A bootstrap test based on the
KS statistic
shows the $r$ values are consistent with this
distribution ($p$-value $0.1$).

Because there are relatively few votes by fans, we have a larger variance in estimates of $\Rfan$ than for $\Rnonfan$. In particular, 17 stories have no votes by fans leading to a maximum likelihood estimate $\Rfan=0$, though with a large confidence interval. The remaining values are approximately lognormally distributed with maximum likelihood estimates of the
mean and standard deviation of $\log(\Rfan)$ equal to $-1.8\pm0.1$ and
$0.75\pm0.08$, respectively. The KS statistic indicates the weaker fit, with a $p$-value of $0.04$. Due to the relatively few votes, the discrete nature of the observations likely significantly affects the estimates. For example, a story with no fans among the early votes may reflect a submitter with no fans and a low, but nonzero, interestingness for fans. A subsequent vote by a highly connected user would expose the story to many fans, possibly leading to many votes that the model would miss by assuming $\Rfan=0$. One approach to this difficulty is using the lognormal distributions of $r$ values as priors in the estimation. This procedure somewhat improves performance, as discussed below.

\begin{figure}[tb]
\centering    \begin{tabular}{c}
      \includegraphics[width=\figwidth]{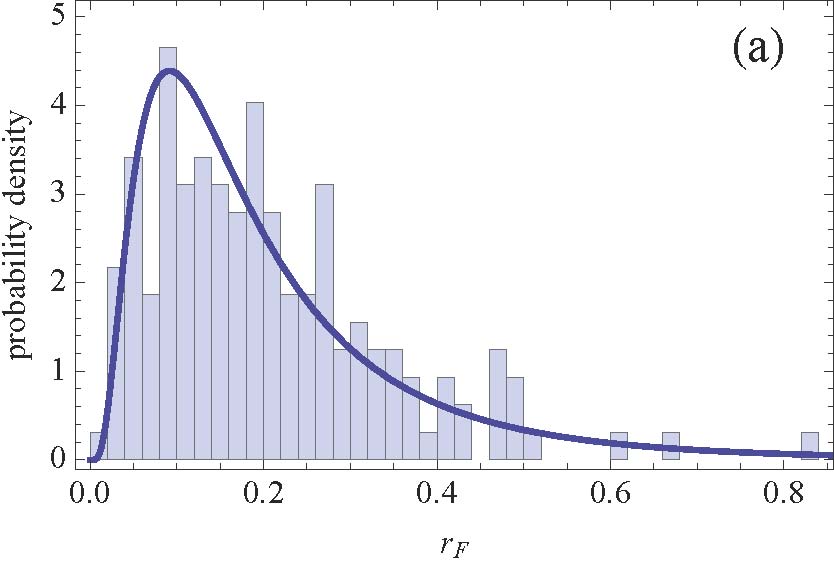} \\
      \includegraphics[width=\figwidth]{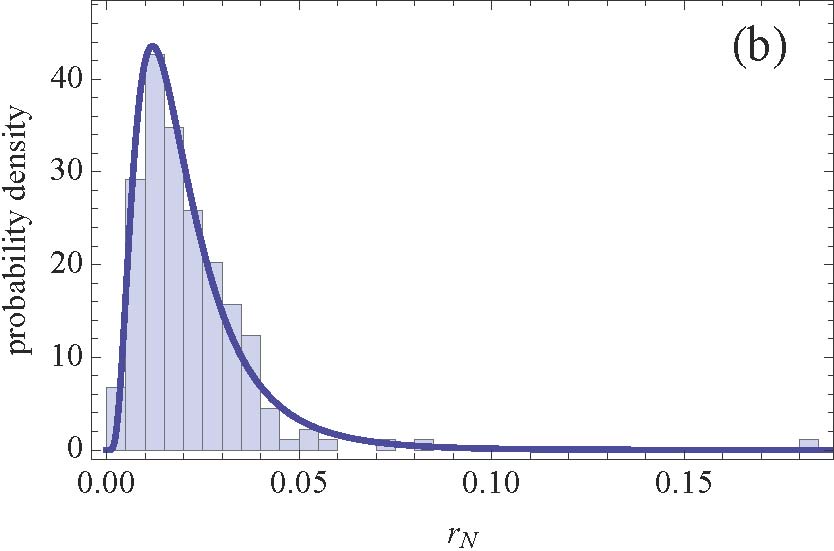}\\
  \end{tabular}
\caption{Distribution of interestingness for (a) fans, and (b) non-fans. The curves are  lognormal fits to the values. Note the different ranges for the horizontal scales in the two plots: $\Rfan$ values tend to be significantly larger than $\Rnonfan$ values.} \figlabel{r values}
\end{figure}

\begin{figure}[t]
\centering
\includegraphics[width=\figwidth]{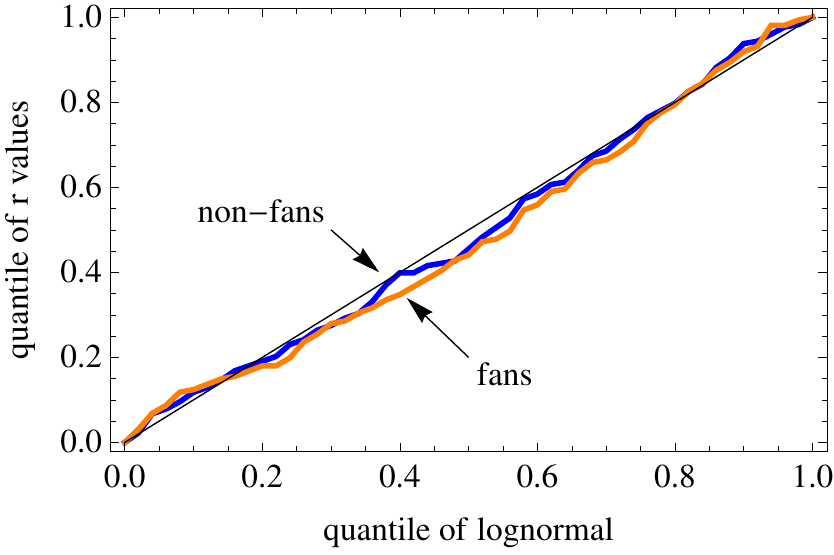}
\caption{Quantile-quantile plot comparing the observed distribution for $\Rfan$ (fans) and $\Rnonfan$ (non-fans) with the corresponding lognormal distribution fits (thick curves). For comparison, the thin straight line from 0 to 1 corresponds to a perfect match between the data and the distribution.}\figlabel{r value quantiles}
\end{figure}

Overall, \fig{rF vs rN} shows there is little relation between how interesting a story is to fans and other users: the correlation between  $\Rfan$ and $\Rnonfan$ is $-0.11$. A randomization test indicates this small correlation is only marginally significant, with $p$-value $0.05$ of arising from uncorrelated values.
The relationship between interestingness for fans and other users indicates a considerable variation in how widely stories appeal to the general user community. Specifically, the ratio $\Rfan/\Rnonfan$ ranges from 0 to 87, with median $9.3$.
The high values correspond to stories that do not get a large number of votes, indicating they are of significantly more interest to the fans of voters than to the general user population, i.e., ``niche interest'' stories (corresponding to the upper left points in \fig{rF vs rN}). As described below, this observation is useful to improve prediction of how popular a story will become based on reaction of early voters. Identifying niche interest stories could also aid user interface design by selectively highlighting stories on the friends interface that have particularly large estimated values of $\Rfan$.
Stories with high ratios of $\Rfan/\Rnonfan$ tend to be promoted after fewer votes than those stories with low ratios.

\begin{figure}[t]
\centering
\includegraphics[width=\figwidth]{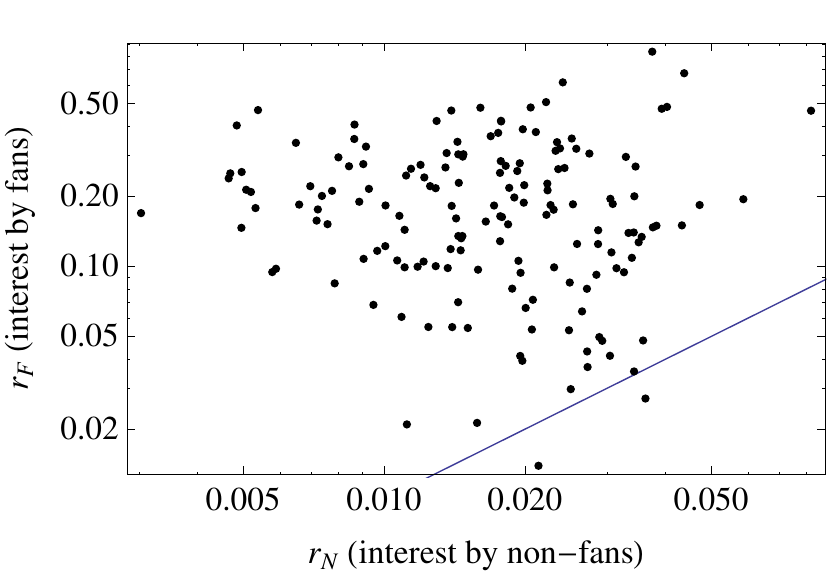}
\caption{Log-log plot comparing estimated interestingness to fans ($\Rfan$) and non-fans ($\Rnonfan$) for 161 promoted stories with votes from fans (so the estimate of $\Rfan$ is positive). All the stories in our data set had non-fan votes, giving all the estimates for $\Rnonfan$ as positive numbers. The line indicates where $\Rfan=\Rnonfan$.}\figlabel{rF vs rN}
\end{figure}

An earlier study~\cite{Lerman08wosn} noted a curious phenomenon: namely, stories that initially spread quickly through the network, i.e., receive a large proportion of early votes from fans, end up not becoming very popular; vice versa, stories that initially spread slowly through the fan network end up becoming popular. This phenomenon appears to be a generic feature of information diffusion on social networks and has also been observed on blog networks~\cite{Colbaugh10isi} and in Second Life~\cite{bakshy09}.

\begin{figure}[t]
\centering
\includegraphics[width=\figwidth]{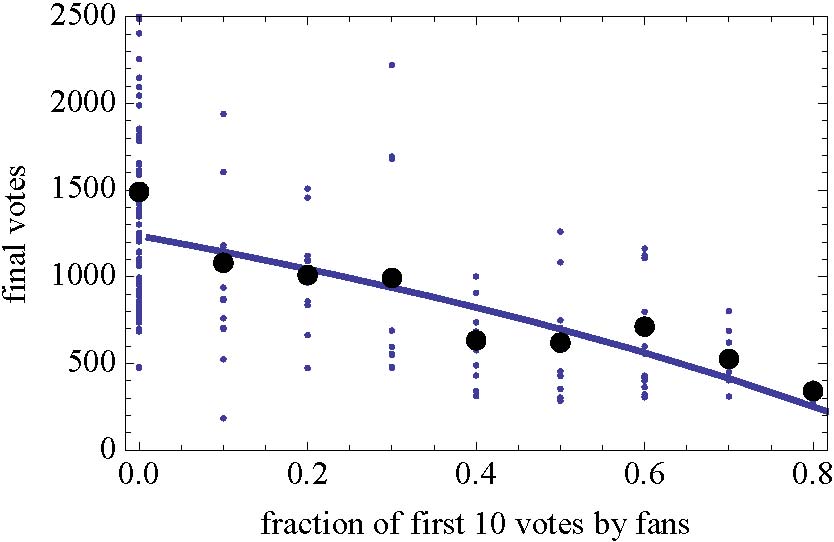}
\caption{Relation between final number of votes and the fraction of votes by fans among a story's first 10 votes. Small points are individual stories and the large points are the mean values for each number of votes by fans. The curve shows the model prediction.
}\figlabel{final vs early fan votes}
\end{figure}

\fig{final vs early fan votes} shows that our model explains this relationship, which arises from the difference in interestingness for fans and non-fans. Specifically, a low fraction of early votes by fans indicates $\Rnonfan$ is relatively large to produce the early non-fan votes in spite of the lower visibility of upcoming stories to non-fan users. Once the story is promoted, it then receives relatively more votes from the general user community (most of whom are not fans of prior voters).  The separation of effects of visibility and interestingness with our model improves this discrimination compared to just using the raw number of votes by fans and non-fans without regard for the story visibility at the time of the votes.
For example, the correlation between the final number of votes and $\Rnonfan/\Rfan$ is $0.72$ compared to $0.64$ for the correlation between the final number of votes and $\nonfanVotes/\fanVotes$.

\subsection{Discussion}

This model with niche interests captures the consequences of link choices: people tend to become fans of users who submit or vote on stories of interest to that person. The ease of incorporating such additional detail is a useful feature of stochastic models.

Comparing the two models illustrates the practical challenges of incomplete or limited data. For example, data scraped from a web site can have errors due to unusual user names or unanticipated characters in story titles. Even when web sites provide an interface to collect data (as Digg provided after the data used in this paper was extracted), subtle differences in interpretation of the data fields can still arise, as when users who no longer have Digg account are all given the same name ``inactive'' and hence appear to be the same user if not specifically checked for in the script collecting the data.

In particular, the ``law of surfing'' parameter estimates for the two models are significantly different, a consequence of the log-likelihood being a fairly flat function of these parameters. This arises due to the relatively weak constraints that vote history provides on \emph{views}, i.e., how many pages of upcoming or front page stories users choose to view during a visit to Digg. For stronger constraints on this behavior, data would ideally include the pages users actually viewed. While such data is in principle available via the web site access logs, this information is not publicly available for Digg. Similarly, the promotion algorithm used by Digg is deliberately not made public to reduce the potential for story submitters to game the system. To the extent that model parameter estimates differ from those that would be possible with this additional data, the stochastic approach identifies potential advantages models can provide the web site provider, with access to more precise data on user behavior, e.g., for predicting popularity of newly submitted stories. More generally, the sensitivity of parameters to the available measured data can suggest additional aspects of user behavior that would be most useful to determine, leading to more focus in future data collection and instrumentation of web communities. Alternatively, when the models indicate several different types of data could provide the required information, selecting the types most acceptable to the user community (e.g., privacy preserving) can facilitate the data collection while providing opportunities for more accurate models to guide the development of the web site and its usefulness to its community.

A related data quality issue is the length of time over which data is collected. On the one hand, collecting data for long periods can improve model parameter estimation by providing many more samples. On the other hand, web sites often rearrange or add features to their user interfaces, which change how users find content. Digg also occasionally changes the promotion algorithm. That is, the stochastic behavior associated with the site is nonstationary rather than arising from a fixed distribution. Moreover, over longer periods of time new users join the site and some users become inactive. Thus one can't simply improve the model parameter estimation by collecting data over longer periods of time~\cite{hogg09c, wilkinson08}. Instead, the models must be extended to include these additional time-dependent behaviors.

In addition to improving quantitative estimation, similar qualitative behaviors seen with different models identify areas for further investigation. For example, in the two models presented here, the distribution of interestingness over the stories shows a lognormal distribution. This suggests there is an underlying multiplicative process giving rise to the observed values~\cite{redner90, mitzenmacher04}. Specifically, the lognormal distribution arises from the multiplication of random variables in the same way that the central limit theorem leads to the normal distribution from the addition of random variables under weak restrictions on their variance and correlations. Thus an important question raised by these models is identifying the story characteristics and user behaviors that combine multiplicatively to lead to the observed lognormal distributions. Identifying such properties would give a more detailed understanding of what leads to interesting content independent of the effects of visibility provided by the web site.

\section{Model-based Prediction}
\label{sec:prediction}

As discussed above, predicting popularity in social media from intrinsic properties of newly submitted content is difficult~\cite{salganik06}. However, users' early reactions provide some measure of predictability~\cite{hogg09c,Kaltenbrunner07,Lerman08wosn,szabo09,Lerman10www}.
The early votes on a story allow us to estimate its interestingness to fans and other users. We can then use the model to predict how the story will accumulate additional votes. These predictions are for expected values and cannot account for the large variation due, for example, to a subsequent vote by a highly connected user which leads to a much larger number of votes.

As one prediction example, we evaluate whether a story will receive at least 500 votes. Predicting whether a story will attract a large number of votes, rather than the precise number of votes, is a useful criterion for predicting whether the story will ``go viral'' and become very popular. This is exactly Digg's intention behind using using crowd sourcing to select a subset of submitted content to feature on the front page~\cite{Lerman08wosn}. The 500 vote threshold is a useful rule of thumb, as that is close to the median popularity value in a large sample of Digg stories~\cite{Lerman10icwsm,wu07}.

\tbl{prediction} compares the predictions with different methods, including a constrained version of our model with $\Rfan=\Rnonfan$, which assumes no systematic difference in interest between fans and other users.


We also compare with direct extrapolation from the early votes. In this procedure, with $v$ votes observed at time $t$, we extrapolate to $v \tFinal/t$, where we take $\tFinal$ to be 72 hours, a time by which stories have accumulated all, or nearly all, the votes they will ever get. We use a least squares linear fit between these observed and extrapolated values. A pairwise bootstrap test indicates the model has a lower prediction error than this extrapolation with $p$-value of $10^{-2}$.

This extrapolation method is similar to that used to predict final votes from the early votes~\cite{szabo09}, but with two differences: 1) we extrapolate from the time required for the story to acquire a given number of votes instead of the number of votes at a given time, and 2) we use early votes after submission (i.e., including when the story is upcoming, where the social network has a large effect) instead of early votes after promotion.

\begin{table}
\centering
\begin{tabular}{rccc}
			& \multicolumn{2}{c}{model} &direct  \\
			& distinct $r$ & same $r$ & extrapolation \\ \hline
first 216 votes & 10\% & 12\% & 21\% \\ 
first 10 votes & 18\% & 23\% & 29\% \\ 
\end{tabular}
\caption{Prediction errors on whether a story receives at least 500 votes. The table compares three methods: 1) the full model which allows distinct values for $\Rfan$ and $\Rnonfan$, 2) the model constrained to have $\Rfan=\Rnonfan$, and 3) direct extrapolation from the rate the story accumulates votes. This comparison involves 178 promoted stories, of which 137 receive at least 500 votes.}\tbllabel{prediction}
\end{table}

In the case of prediction based on the first 10 votes, which is before the stories are promoted, an additional question is how well the model predicts whether the story will eventually be promoted. We find a $25\%$ error rate in predicting promotion based on the first 10 votes.

We can improve predictions from early votes by using the lognormal distributions of $\Rfan$ and $\Rnonfan$, shown in \fig{r values}, as the prior probability to combine with the likelihood from the observations according to Bayes theorem.
Specifically, instead of maximizing the likelihood of the observed votes, $P(r|\mbox{votes})$, as discussed above, this approaches maximizes the posterior probability, which is proportional to $P(r|\mbox{votes}) \Pprior(r)$ where $\Pprior$ is taken to be the lognormal distribution $\Plognormal$ in \eq{lognormal} with parameters from the fits shown in \fig{r values}.

\begin{figure}[t]
\centering
\includegraphics[width=\figwidth]{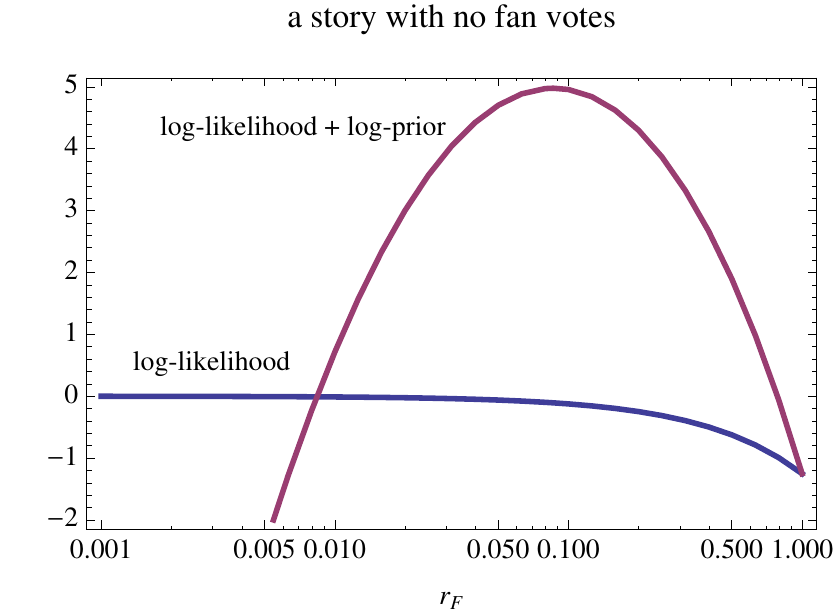}
\caption{Comparison of log-likelihood (i.e., $\log P(r|\mbox{votes})$) and log-likelihood plus $\log (\Pprior(r))$ for estimating $\Rfan$ for a story with no fan votes. The maximum of the log-likelihood is at $\Rfan=0$ while the maximum with the prior is $\Rfan=0.086$.} \label{fig.rF with prior}
\end{figure}

This method gives little change in estimates of $\Rnonfan$, due to the relatively large number of non-fan votes on each story. However, using the prior makes large changes in some of the $\Rfan$ estimates, thereby avoiding the small number of extreme predictions made by poor estimates. Using this prior to aid estimation is particularly significant when there are no votes by fans among the early votes, leading to an estimate of $\Rfan=0$, but later a user with many fans votes on the story. In this case, as illustrated in \fig{rF with prior}, using the lognormal as a prior gives a positive estimate for $\Rfan$, thereby predicting some votes by any subsequent users who are fans of earlier voters.

By avoiding these extreme cases, this procedure improves the correlation between predicted and actual final votes as well as the predicted rank ordering of the stories (i.e., whether the story is likely to be relatively popular) as seen with a larger value of the Spearman rank correlation when using the prior distribution. For example, when predicting based on the first 10 votes, using this prior increases the Spearman rank correlation between predicted and actual number of votes from $0.46$ to $0.53$. For comparison, this correlation for direct extrapolation from the first 10 votes is $0.32$ and is $0.34$ for the model constrained to have $\Rfan=\Rnonfan$. Pairwise bootstrap tests indicate the differences between these values are significant with $p$-values less than $10^{-3}$, except the difference between the last two cases has $p$-value of $10^{-2}$.

\section{Related work}
\label{sec:related}
The Social Web provides massive quantities of available data about the behavior of large groups of people. Researchers are using this data to study a variety of topics, including detecting~\cite{Adamic04,Leskovec07kdd} and influencing~\cite{Domingos01,Kempe03} trends in public opinion, and dynamics of information flow in groups~\cite{Wu04,Leskovec07}.

Several researchers examined the role of social dynamics in explaining and predicting distribution of popularity of online content.
Wilkinson~\cite{wilkinson08} found broad distributions of popularity and user activity on many social media sites and showed that these distributions can arise from simple macroscopic dynamical rules.
Wu \& Huberman~\cite{wu07} constructed a phenomenological model of the dynamics of collective attention on Digg. Their model is parameterized by a single variable that characterizes the rate of decay of interest in a news article. Rather than characterize evolution of votes received by a single story, they show the model describes the distribution of final votes received by promoted stories. Our model offers an alternative explanation for the distribution of votes. Rather than novelty decay, we argue that the distribution can also be explained by the combination of a non-uniform variations in the stories' inherent interest to users and effects of user interface, specifically decay in visibility as the story moves to subsequent front pages. Such a mechanism can also explain the distribution of popularity of photos on Flickr, which would be difficult to characterize by novelty decay.
Crane \& Sornette~\cite{CraneSornette08} analyzed a large number of videos posted on You\-Tube and found that collective dynamics was linked to the inherent quality of videos. By looking at how the observed number of votes received by videos changed in time, they could separate high quality videos, whether they were selected by You\-Tube editors or spontaneously became popular, from junk videos. This study is similar in spirit to our own in exploiting the link between observed popularity and content quality. However, while this, and Wu \& Huberman study, aggregated data from tens of thousands of individuals, our method focuses instead on the \emph{microscopic} dynamics, modeling how individual behavior contributes to the observed popularity of content. In \cite{Lerman10www} we used the simple model of social dynamics, reviewed in this paper, to predict whether Digg stories will become popular. The current paper improves on that work.

Researchers found statistically significant correlation between early and late popularity of content on Slashdot~\cite{Kaltenbrunner07}, Digg and You\-Tube~\cite{szabo09}. Specifically, similar to our study, Szabo \& Huberman~\cite{szabo09} predicted long-term popularity of stories on Digg. Through large-scale statistical study of stories promoted to the front page, they were able to predict stories' popularity after 30 days based on their popularity one hour after promotion. Unlike our work, their study did not specify a mechanism for evolution of popularity, and simply exploited the correlation between early and late story popularity to make the prediction. Our work also differs in that we predict popularity of stories shortly after submission, long before they are promoted.
Several researchers~\cite{Lerman08wosn,bakshy09,Colbaugh10isi}  found that early diffusion of information across an interlinked community is a useful predictor of how far it will spread across the network in general. Both \cite{Lerman08wosn} and \cite{Colbaugh10isi} exploited the anti-correlation between these phenomena to predict final popularity. Specifically, the former work used anti-correlation between the number of early fan votes and stories' eventual popularity on Digg
to predict whether stories submitted by well connected users will become popular. That work exploited social influence only to make the prediction, and the results were not applicable to stories submitted by poorly connected users which were not quickly discovered by highly connected users. In contrast, the approach described in this paper considers effects of social influence regardless of the connectedness of the submitter, and also accounts for story quality in making a prediction about story popularity. An interesting open question is the nature of the social influence on voting. In our model, the influence has two components: increased visibility of a story to fans due to the friends interface and the higher interestingness of the story to fans. This higher interestingness could be due to self-selection, whereby users become fans of people whose submissions or votes are of particular interest. Alternatively, users could be directly influenced by the activities of others~\cite{salganik06}, with the possibility that this influence depends not just on whether friends vote on a story but also how many friends do so~\cite{centola10}.

\section{Conclusion}
In the vast stream of new user-generated content, only a few items will prove to be popular, attracting a lion's share of attention, while the rest languish in obscurity. Predicting which items will become popular is exceedingly difficult, even for people with significant expertise. This prediction difficulty arises because popularity is weakly related to inherent content quality and social influence leads to an uneven distribution of popularity that is sensitive to the early choices of users in the social network. We described how stochastic models of user behavior on a social media web site can partially address this prediction challenge by quantitatively characterizing evolution of popularity. The model shows how popularity is affected by item quality and social influence. We evaluated the usefulness of this approach for the social news aggregator Digg, which allows users to submit and vote on news stories. The number of votes a story accumulates on Digg shows its popularity. In earlier work we developed a model of social voting on Digg, which describes how the number of votes received by a story changes in time. In that model, knowing how interesting a story is to the user community, on average, and how connected the submitter is fully determines the evolution of the story's votes. This leads to an insight that a model can be used to predict story's popularity from the initial reaction of users to it. Specifically, we use observations of evolution of the number of votes received by a story shortly after submission to estimate how interesting it is, and then use the model to predict how many votes the story will get after a period of a few days. Model-based prediction outperforms other methods that exploit social influence only, or correlation between early and late votes received by stories. We improved prediction by developing a more fine-grained model that differentiates between how interesting a story is to fans and to the general population.


\bibliography{references}
\bibliographystyle{plain}

\end{document}